\documentclass[twocolumn]{aastex62}

\usepackage{amsmath}
\usepackage{graphicx}   
\usepackage{bm}         
\usepackage{natbib}
\usepackage{etoolbox}
\usepackage{rotating}
\usepackage{array}
\usepackage{booktabs}
\usepackage{amssymb}
\usepackage[T1]{fontenc}

\makeatletter
\patchcmd{\NAT@citex}
  {\@citea\NAT@hyper@{%
     \NAT@nmfmt{\NAT@nm}%
\hyper@natlinkbreak{\NAT@aysep\NAT@spacechar}{\@citeb\@extra@b@citeb}%
     \NAT@date}}
  {\@citea\NAT@nmfmt{\NAT@nm}%
   \NAT@aysep\NAT@spacechar\NAT@hyper@{\NAT@date}}{}{}

\patchcmd{\NAT@citex}
  {\@citea\NAT@hyper@{%
     \NAT@nmfmt{\NAT@nm}%
\hyper@natlinkbreak{\NAT@spacechar\NAT@@open\if*#1*\else#1\NAT@spacechar\fi}%
       {\@citeb\@extra@b@citeb}%
     \NAT@date}}
  {\@citea\NAT@nmfmt{\NAT@nm}%
\NAT@spacechar\NAT@@open\if*#1*\else#1\NAT@spacechar\fi\NAT@hyper@{\NAT@date}}
  {}{}
\makeatother

\makeatletter
\DeclareRobustCommand{\textsupsub}[2]{{%
  \m@th\ensuremath{%
    ^{\mbox{\fontsize\sf@size\z@#1}}%
    _{\mbox{\fontsize\sf@size\z@#2}}%
  }%
}}
\makeatother

\newcommand{\lsun}{\mbox{\,$L_\odot$}}
\newcommand{\msun}{\mbox{\,$M_\odot$}}

\newcommand{\msunyr}{\mbox{\,$M_\odot$ yr$^{-1}$}}
\newcommand{\mdotstar}{\mbox{$\dot{M}_{\star}$}}
\newcommand{\kms}{\mbox{\,km\,s$^{-1}$}}

\newcommand{\ee}[1]{\mbox{${} \times 10^{#1}$}}
\newcommand{\eten}[1]{\mbox{$10^{#1}$}}
\newcommand{\jj}[2]{\mbox{$J = #1\rightarrow#2$}}

\newcommand{\cc}{\mbox{cm$^{-3}$}}

\newcommand{\vlsr}{\mbox{$v_\text{lsr}$}}

\newcommand{\source}{L1551 IRS~5}
\newcommand{\sofia}{SOFIA}
\newcommand{\oi}{[O~\textsc{i}]}
\newcommand{\cii}{[C~\textsc{ii}]}
\newcommand{\feii}{[Fe~\textsc{ii}]}

\newcommand{\water}{\mbox{H$_{2}$O}}

\shorttitle{[OI] Outflow in \source}
\shortauthors{Yang et al.}

\begin{document}

\title{Atomic Shocks in the Outflow of \source\ Identified with SOFIA-upGREAT Observations of \oi}

\author[0000-0001-8227-2816]{Yao-Lun Yang}
\affiliation{Department of Astronomy, University of Virginia, Charlottesville, VA 22904, USA}

\author[0000-0001-5175-1777]{Neal J. Evans II}
\affiliation{Department of Astronomy, The University of Texas at Austin, Austin, TX 78712, USA}

\author[0000-0001-8913-925X]{Agata Karska}
\affiliation{Institute of Astronomy, Faculty of Physics, Astronomy and Informatics, Nicolaus Copernicus University, Grudziadzka 5, 87-100 Torun, Poland}

\author[0000-0003-1159-3721]{Lars E. Kristensen}
\affiliation{Niels Bohr Institute \& Centre for Star and Planet Formation, Copenhagen University, {\O}ster Voldgade 5--7, 1350 Copenhagen K, Denmark}

\author[0000-0002-1316-1343]{Rebeca Aladro}
\affiliation{Max-Planck-Institut f\"{u}r Radioastronomie, Auf dem H\"{u}gel 69, D-53121 Bonn, Germany}

\author[0000-0002-3835-3990]{Jon P. Ramsey}
\affiliation{Department of Astronomy, University of Virginia, Charlottesville, VA 22904, USA}

\author[0000-0003-1665-5709]{Joel D. Green}
\affiliation{Space Telescope Science Institute, Baltimore, MD 02138, USA}
\affiliation{Department of Astronomy, The University of Texas at Austin, Austin, TX 78712, USA}

\author[0000-0003-3119-2087]{Jeong-Eun Lee}
\affiliation{Department of Astronomy \&\ Space Science, Kyung Hee University, Gyeonggi 446-701, Korea  \\
School of Space Research, Kyung Hee University, Yongin-shi, Kyungki-do 449-701, Korea}

\correspondingauthor{Yao-Lun Yang}
\email{yaolunyang.astro@gmail.com}

\begin{abstract}
We present velocity resolved SOFIA/upGREAT observations of \oi\ and \cii\ lines toward a Class I protostar, \source, and its outflows.  The SOFIA observations detect \oi\ emission toward only the protostar and \cii\ emission toward the protostar and the red-shifted outflow.  The \oi\ emission has a width of $\sim$100 \kms\ only in the blue-shifted velocity, suggesting an origin in shocked gas.  The \cii\ lines are narrow, consistent with an origin in a photodissociation region.  Differential dust extinction from the envelope due to the inclination of the outflows is the most likely cause of the missing red-shifted \oi\ emission.  Fitting the \oi\ line profile with two Gaussian components, we find one component at the source velocity with a width of $\sim$20 \kms\ and another extremely broad component at $-30$ \kms\ with a width of 87.5 \kms, the latter of which has not been seen in \source.  The kinematics of these two components resemble cavity shocks in molecular outflows and spot shocks in jets.  Radiative transfer calculations of the \oi, high-$J$ CO, and \water\ lines in the cavity shocks indicate that \oi\ dominates the oxygen budget, making up more than 70\% of the total gaseous oxygen abundance and suggesting [O]/[H] of $\sim$1.5\ee{-4}. Attributing the extremely broad \oi\ component to atomic winds, we estimate the intrinsic mass loss rate of (1.3$\pm$0.8)\ee{-6} \msunyr.  The intrinsic mass loss rates derived from low-$J$ CO, \oi, and HI are similar, supporting the model of momentum-conserving outflows, where the atomic wind carries most momentum and drives the molecular outflows.
\end{abstract}

\section{Introduction}

Protostellar outflows are the primary mechanism of angular momentum removal that facilitates the growth of protostars.  Outflows also allow transportation of materials from the disk to the protostars, injecting kinetic energy into the surroundings.  The outflow-envelope interaction can lead to UV-irradiated outflow cavities and shocks both along the outflows and on the cavity walls \citep[e.g., ][]{2012AA...542A...8K,2014AA...572A...9K,2018ApJ...860..174Y}. Atomic oxygen fine-structure lines are important cooling lines that trace shocked gas in outflows \citep[e.g.,][]{2015ApJ...801..121N,2018ApJS..235...30K}, especially the $^3P_1\rightarrow{^3}P_2$ transition at 63 \micron.  If the \oi\ emission comes from dissociative J-shocks in the outflow, where \oi\ dominates the cooling, the \oi\ luminosity could trace the mass loss rate \citep{1985Icar...61...36H,1989ApJ...342..306H}.  Studies using observations of the Herschel Space Observatory surveyed the \oi\ 63 \micron\ line from low-mass to high-mass protostars with many results suggesting that the line originates in UV-irradiated shocks  \citep{2014AA...562A..45K,2015ApJ...801..121N,2018ApJS..235...30K}. However, the velocity resolution of the Herschel instruments was not sufficient to resolve the line profile of \oi, although significant velocity offsets of the line centroid were seen in several outflows \citep{2010AA...518L.121V,2014ApJS..214...21L,2016AA...594A..59R}.

With the advent of the Stratospheric Observatory for Infrared Astronomy (SOFIA), it became possible to obtain velocity-resolved spectra of the \oi\ line.  In a massive star-forming region, \citet{2015AA...584A..70L} show complex line profiles of \oi\ observed by SOFIA, including high-velocity line wings and absorption around the source velocity.  Thus, a faithful representation of different components in the outflow requires velocity-resolved line profiles.  Toward NGC 1333 IRAS 4A, a low-mass protostar, \citet{2017AA...601L...4K} used SOFIA to observe the \oi\ 63 \micron\ line in the R1 shock knot along the outflow.  The line profile of the \oi\ emission is similar to that of CO\,\jj{16}{15} and shock-excited \water\ lines, implying that they trace the same shock component.  They suggest that CO is the major carrier of volatile oxygen in the shock and only $\sim15$\%\ of the oxygen is atomic.  
Similar high-spectral-resolution observations of \oi\ have been only performed in a few intermediate-mass \citep{2017AA...602A...8G,2021AA...653A.108S} and high-mass protostars \citep{2015AA...584A..70L,2018AA...617A..45S} due to the lower \oi\ brightness in low-mass protostars.

\source\ has an exceptionally high \oi\ 63 \micron\ luminosity among low-mass protostars \citep{2016AJ....151...75G}, making it an ideal target to study the kinematics and energetics of \oi\ in low-mass protostars.  \source\ is a Class I binary protostar \citep{1998Natur.395..355R} at a distance of $147.3\pm0.5$ pc \citep{2018ApJ...859...33G}.  The two protostars are separated by $\sim$50 au (0\farcs{3}) and each hosts its own disk with a radius of $\sim$10 au \citep{1998Natur.395..355R,2006ApJ...653..425L,2016ApJ...826..153L}, sharing a circumbinary disk  \citep{2019ApJ...882L...4C}.  Each source drives its own jet observed in free-free emission \citep{2003ApJ...586L.137R} and the \feii\ 1.644 \micron\ line \citep{2005ApJ...618..817P,2009ApJ...694..654P}.  The \feii\ emission is only observed at blue-shifted velocities, extending $\sim$10\arcsec\ to the southwest with a maximum velocity of $-400$ \kms.  \source\ has molecular bipolar outflows in the northeast-southwest direction, first recognized by \citet{1980ApJ...239L..17S} with CO, which is in fact the first discovery of molecular outflows in protostars.  \citet{2009ApJ...698..184W} constrained a half outflow opening angle of 22\arcdeg\ and a position angle of 57\arcdeg\ using the CO\,\jj{2}{1} line observed by the Submillimeter Array.  \citet{2014ApJ...796...70C} further constrained the disk inclination angle of 60\arcdeg$\pm$5\arcdeg\ with respect to the plane of sky by modeling the Keplerian circumbinary disk, suggesting an inclination of 30\arcdeg\ for the outflow.  \citet{2002AA...382..573F} derived a source velocity of 6.5 \kms\ for \source, while \citet{2013AA...556A..89Y} found a slightly lower source velocity of 6.2 \kms, which has since been adopted by several studies of high-$J$ CO and water emission \citep[e.g., ][]{2017AA...605A..93K}.  Thus, we adopt a source velocity 6.2 \kms\ in this study to ensure consistent comparisons.  Taken together, the long history of studies on \source\ make it one of the best-studied Class I outflows, and thus an ideal target for a more detailed study of \oi\ emission in protostellar outflows.

In this study, we present velocity-resolved spectra of the \oi\ 63 \micron\ and \cii\ 158 \micron\ lines observed towards \source\ with SOFIA.  Section\,\ref{sec:obs} describes the observations and data reduction.  Section\,\ref{sec:results} presents the spectra of \oi\ and \cii\ and compares the observations with archival Herschel data.  Section\,\ref{sec:discussion} discusses the origin of the \oi\ emission in \source, how it related to shocks in the outflow, the oxygen budget in shocks, and the outflow mass loss rate inferred from the \oi\ emission.  Section\,\ref{sec:conclusions} presents the conclusions of this study.

\section{Observations}
\label{sec:obs}
The observations of \source\ (Project ID: 06\_0104, PI: Y.-L. Yang) were conducted with \sofia/\ upGREAT \citep[upgrade German REceiver for Astronomy at Terahertz Frequencies;][]{2016AA...595A..34R,2018JAI.....740014R,doi:10.1142/S2251171718400111} in Cycle 6 (2018 December 5) and 7 (2020 March 5, 10, and 13) with aircraft altitudes ranging from 39,000 to 41,000 feet.  Three positions toward \source\ were observed, including the central protostar as well as the known blue- and red-shifted outflows.  The ``blue'' and ``red'' pointings are separated from the central protostar by 6\farcs{9} along the outflow direction (PA$=$57$^\circ$; \citealt{2009ApJ...698..184W}). 

We used multipixel receivers HFA (high frequency array) and LFA (low frequency array) connected to the Fast-Fourier-Transform spectrometer XFFTS.  The LFA receiver consists of 2$\times$7 pixels for orthogonal polarizations in a hexagonal configuration with a central pixel, while the HFA has the same array geometry but only a single linear polarization.  We configured the HFA to target the \oi\ $^3P_1\rightarrow{^3}P_2$ transition at 4.7447775 THz (63.18 \micron) and the LFA to simultaneously observe the \cii\ $^2P_{3/2}\rightarrow{^2}P_{1/2}$ transition at 1.9005369 THz (157.74 \micron) and the OH doublet $^2\Pi^{-}_{1/2,\,3/2}\rightarrow{^2}\Pi^{+}_{1/2,\,1/2}$ transition at 1.8347598 and 1.83904685 THz (163.40 and 163.12 \micron) in horizontal and vertical polarizations, respectively.  No OH emission was detected in our observations, consistent with previous Herschel/PACS observations \citep{2016AJ....151...75G}, and thus, in this study, we focus on the \oi\ and \cii\ emission.  The half-power beam widths (HPBW) of the HFA and LFA observations are 6\farcs{3} and 14\farcs{1}, respectively.

\begin{figure*}
  \centering
  \includegraphics[width=\textwidth]{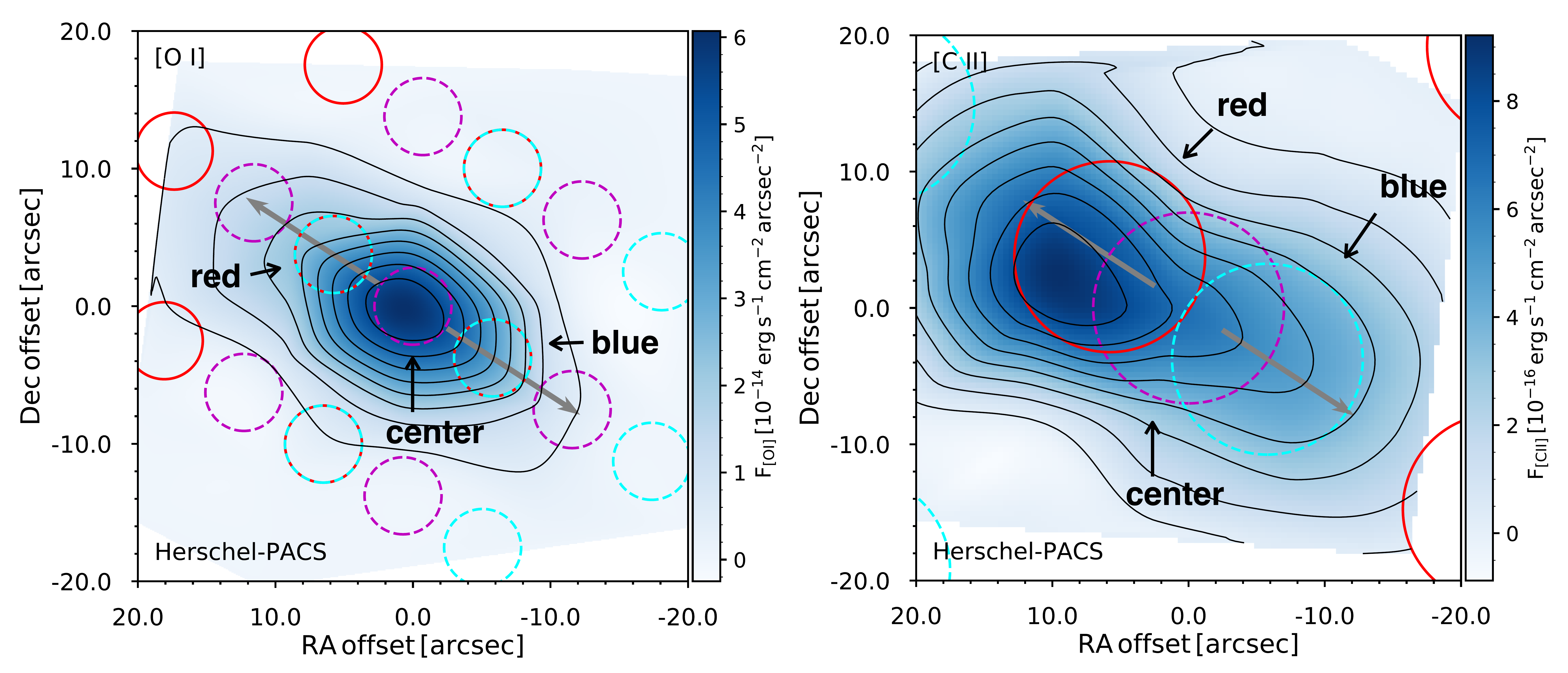}
  \caption{The intensity maps of the \oi\ 63 \micron\ (left) and \cii\ 158 \micron\ (right) lines observed by Herschel-PACS \citep{2014ApJS..214...21L,2016AJ....151...75G}.  The footprints indicate the layouts of HFA and LFA for the \oi\ and \cii\ lines, respectively.  The ``red'', ``center'', and ``blue'' positions are annotated.  The footprints of the HFA and LFA arrays for the ``red'', ``center'', and ``blue'' pointings are shown in red circles, magenta circles, and cyan dashed circles, respectively.  The gray arrows denote the outflow direction (position angle $=$ 57\arcdeg, \citealt{2009ApJ...698..184W}), red-shifted to the northeast and blue-shifted to the southwest.}
  \label{fig:footprint}
\end{figure*}

We rotated the arrays by $-$33$^{\circ}$ to align the center row of three pixels in both HFA and LFA with the outflow direction.  Henceforth, we refer to the positions toward the protostar, the red- and blue-shifted outflows as ``center'', ``red'', and ``blue'' positions (Table\,\ref{tbl:positions} and Figure\,\ref{fig:footprint}).  To maximize the observing efficiency, we utilized the HFA array to simultaneously observe the ``red'' and ``blue'' positions, while observing for the \cii\ line at the ``red'' and ``blue'' positions with individual LFA pointings.  While both HFA and LFA have seven pixels, given the brightness distribution observed by \textit{Herschel}-PACS \citep{2014ApJS..214...21L}, we expected to detect emission only in the three positions chosen.  The observations were performed in Dual Beam Switching mode (DBS).  The offset position was 30\arcsec\ away from \source, perpendicular to the outflow direction, and observed with a chop frequency of 0.626 Hz.  The on-source time for each position and spectral setup is listed in Table\,\ref{tbl:positions}. 

\begin{table*}[htbp!]
  \centering
  \caption{Observed Positions and On-source Times}
  \label{tbl:positions}
  \begin{tabular}{ccccc}
    \toprule
    Name & R.A. (J2000) & Dec. (J2000) & \oi\ on-source time (sec.) & \cii\ \&\ OH on-source time (sec.) \\
    \midrule
    center & 04:31:34.07 & 18:08:04.90 & p50: 132; m50: 165 & 297 \\
    blue & 04:31:33.69 & 18:08:01.10 & p50: 554; m50: 229 & 267 \\
    red & 04:31:34.46 & 18:08:08.70 & p50: 554; m50: n/a & 286 \\
    \bottomrule
  \end{tabular}
\end{table*}

To maximize the bandwidth around the \oi\ line, each position was observed with two spectral set ups centered at $-$50 \kms\ and $+$50 \kms, denoted as ``m50'' and ``p50'', respectively, to cover $-$100 \kms\ to $+$100 \kms\ around the \oi\ line.  The ``red'' position was only observed with the ``p50'' setup due to the cancellation of a scheduled flight.  However, given that only red-shifted emission would be expected at this position anyway, observing only the ``p50'' setup should have captured most of the emission from the red-shifted outflow.  While the HFA was observing the full $-$100 \kms\ to $+$100 \kms\ around the \oi\ line with two spectral setups, the LFA only had one spectral setup; as a result, the LFA spectral setups were observed for twice as long as the HFA setup.

We used \textsc{class} \citep{2013ascl.soft05010G} for data reduction.  All spectra were inspected individually, and those associated with backend or receiver instabilities were dropped and not considered for further reduction.  SOFIA/upGREAT has a native spectral resolution of 244 kHz, corresponding to 0.015 and 0.039 \kms\ resolution at the \oi\ and \cii\ lines.  We first smoothed the spectrum of each observing block to 1 \kms.  Then, we fitted the baseline in the line-free region and subtracted it from each spectrum using a first-order polynomial.  Finally, we averaged all spectra taken with the same setup together, weighted by their respective noise.  The \oi\ observations in the ``m50'' setup were affected by a narrow, but intense, telluric line entering at a velocity of $-40$ \kms\ (in Cycle 7) and $-14$ \kms\ (in Cycle 6). We masked the channels of the atmospheric line in order to remove it from the analysis of the emission.

\section{Results and Analyses}
\label{sec:results}
\subsection{\oi\ 63 ${\it \mu}m$}
\begin{figure}[htbp!]
  \centering
  \includegraphics[width=0.48\textwidth]{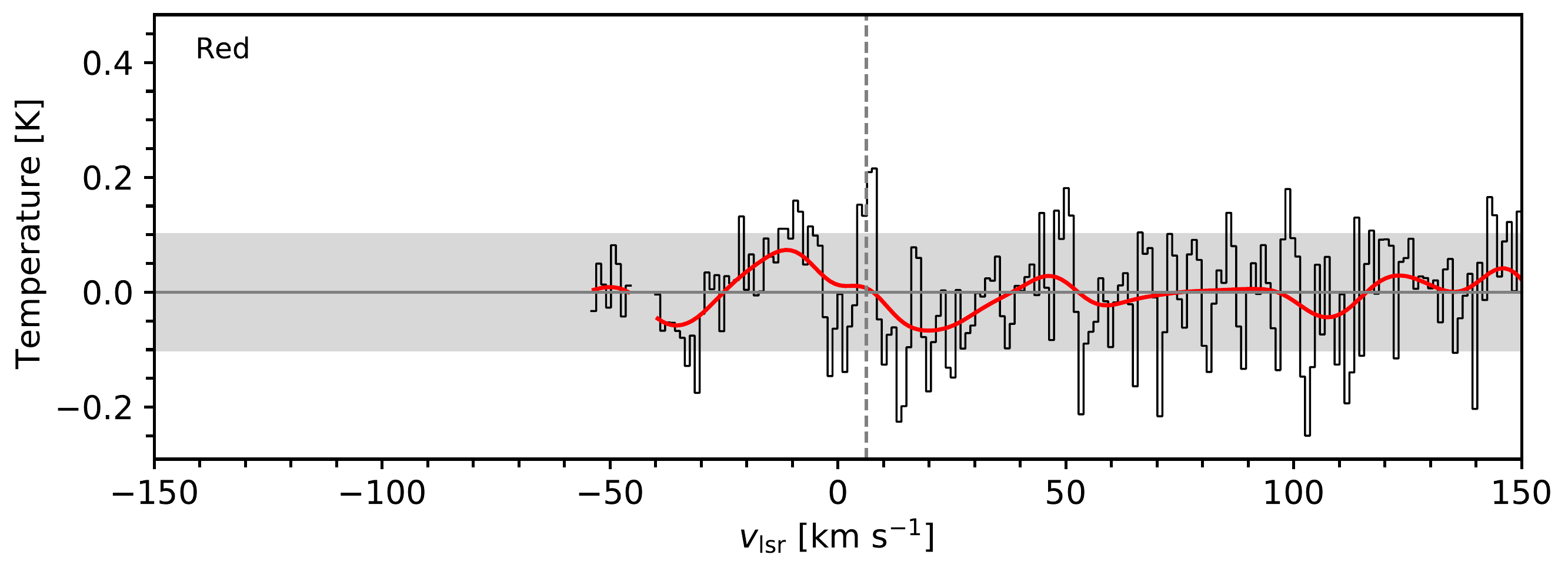}
  \includegraphics[width=0.48\textwidth]{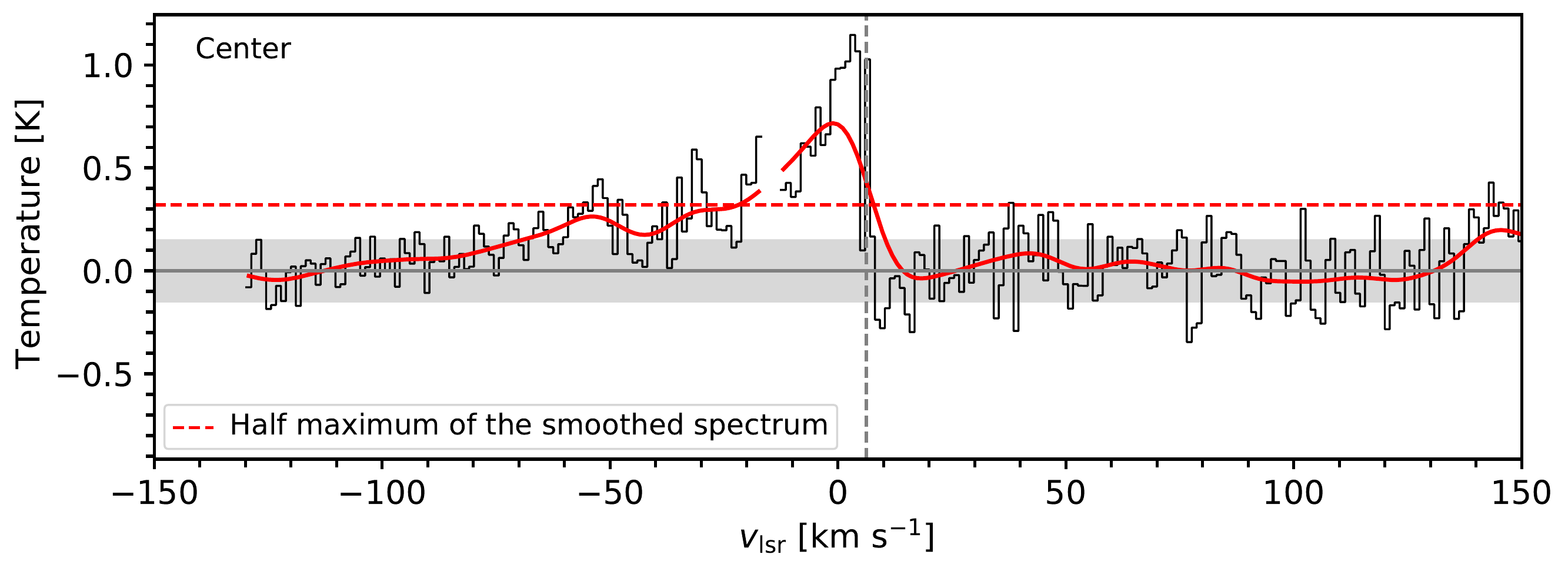}
  \includegraphics[width=0.48\textwidth]{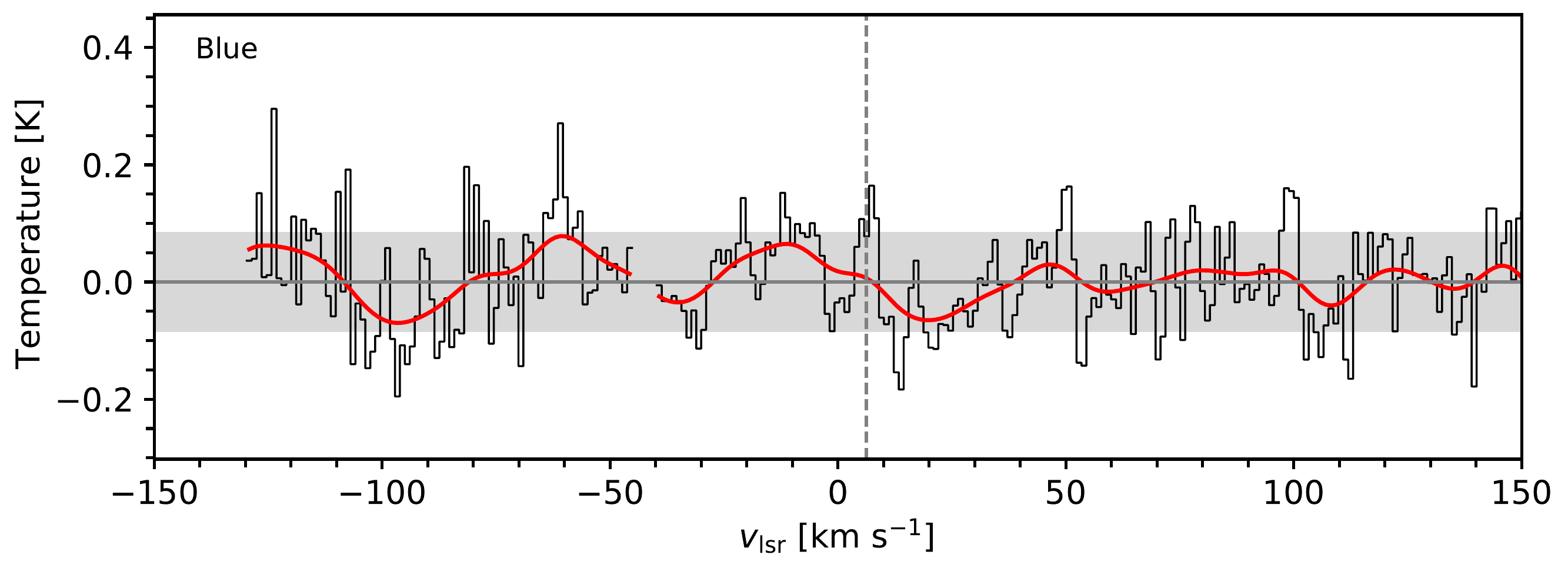}
  \caption{The baseline-subtracted \oi\ spectra toward the ``red,'' ``center, '' and ``blue'' positions from top to bottom.  The black line shows the spectra smoothed to 1 \kms, while the red line shows the spectra smoothed to 5 \kms.  The vertical dashed line indicates the source velocity of 6.2 \kms.  The horizontal red dashed line in the ``center'' spectrum illustrates the half maximum of the spectrum smoothed to 5 \kms.  The shaded region illustrates the $\pm1\sigma$ noise around the baseline.  Channels showing telluric features at $-14$ and $-40$ \kms\ have been masked (see Section\,\ref{sec:obs}).}
  \label{fig:oi}
\end{figure}

To combine two \oi\ spectra taken with two setups of local oscillator frequencies (``m50'' and ``p50''; see Section\,\ref{sec:obs}), we took the weighted average of the two spectra using on-source time as the weight.  Figure\,\ref{fig:oi} shows the \oi\ spectra observed toward the ``red,'' ``center,'' and ``blue'' positions.  For the baseline fitting and subtraction, we selected velocity ranges of $-150\,\kms<\vlsr<-110\,\kms$ and $20\,\kms<\vlsr<130\,\kms$.  The baseline noise is 0.10 K, 0.15 K, and 0.09 K for the ``red,'' ``center, '' and ``blue'' positions, respectively.  The ``center'' spectrum shows a clear detection of the \oi\ 63 \micron\ line, with a peak at around 0--5 \kms\ and a broad blue-shifted tail.  We detect no emission at red-shifted velocities at the ``center'' position. We also do not detect any significant \oi\ emission in the ``red'' or ``blue'' positions in the outflows.  To better characterize the broad blue-shifted feature at the center position, we further smoothed the \oi\ spectra to 5 \kms\ resolution.  At the ``center'' position, the smoothed \oi\ spectrum shows two tentative local peaks at $-$53 \kms\ and $-$31 \kms\ in addition to the prominent peak around the source velocity.  The intensities of these two peaks are roughly equal to the half of the maximum intensity.  If these two peaks belong to the same broad feature, the FWHM of the \oi\ line is then $\sim$100 \kms.  However, if these two peaks are separate features, the primary line profile centered at the source velocity would have a FWHM of $\sim$40 \kms\ along with two lines at $-$53 \kms\ and $-$31 \kms\ with a width of $\sim$10 \kms each.

\subsection{\cii\ 158 ${\it \mu}m$}
\label{sec:result_cii}
\begin{figure}[htbp!]
  \centering
  \includegraphics[width=0.48\textwidth]{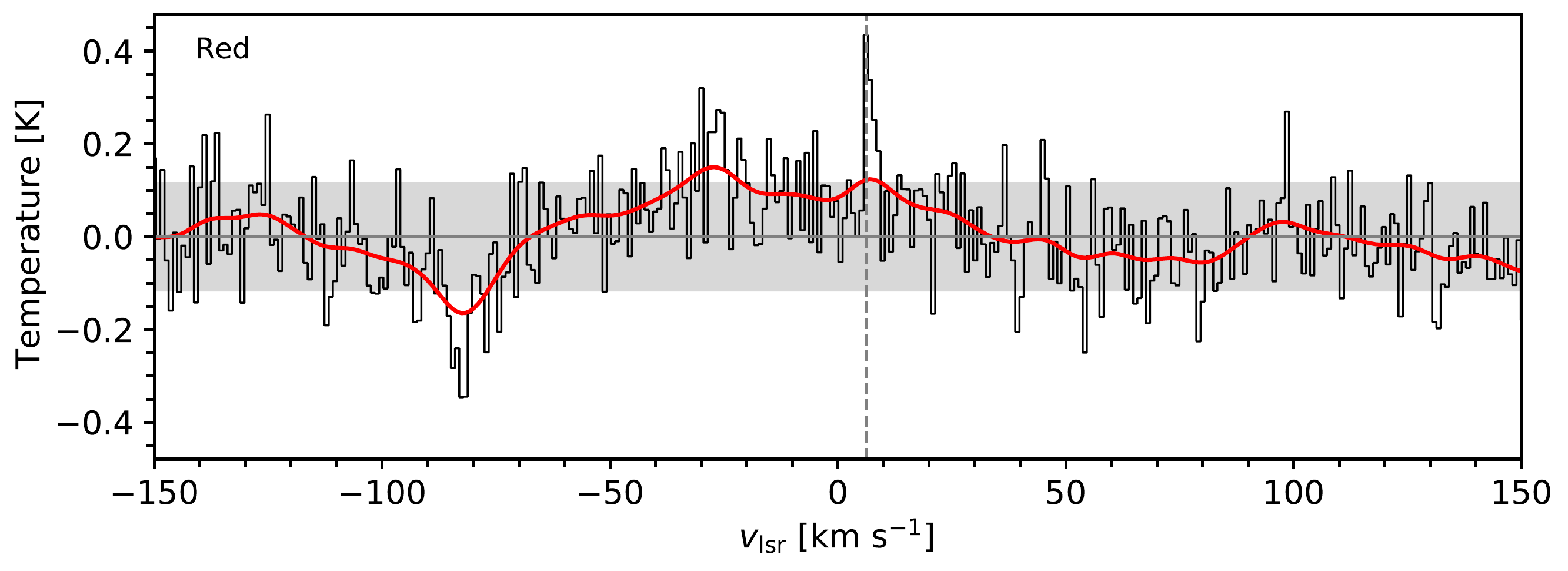}
  \includegraphics[width=0.48\textwidth]{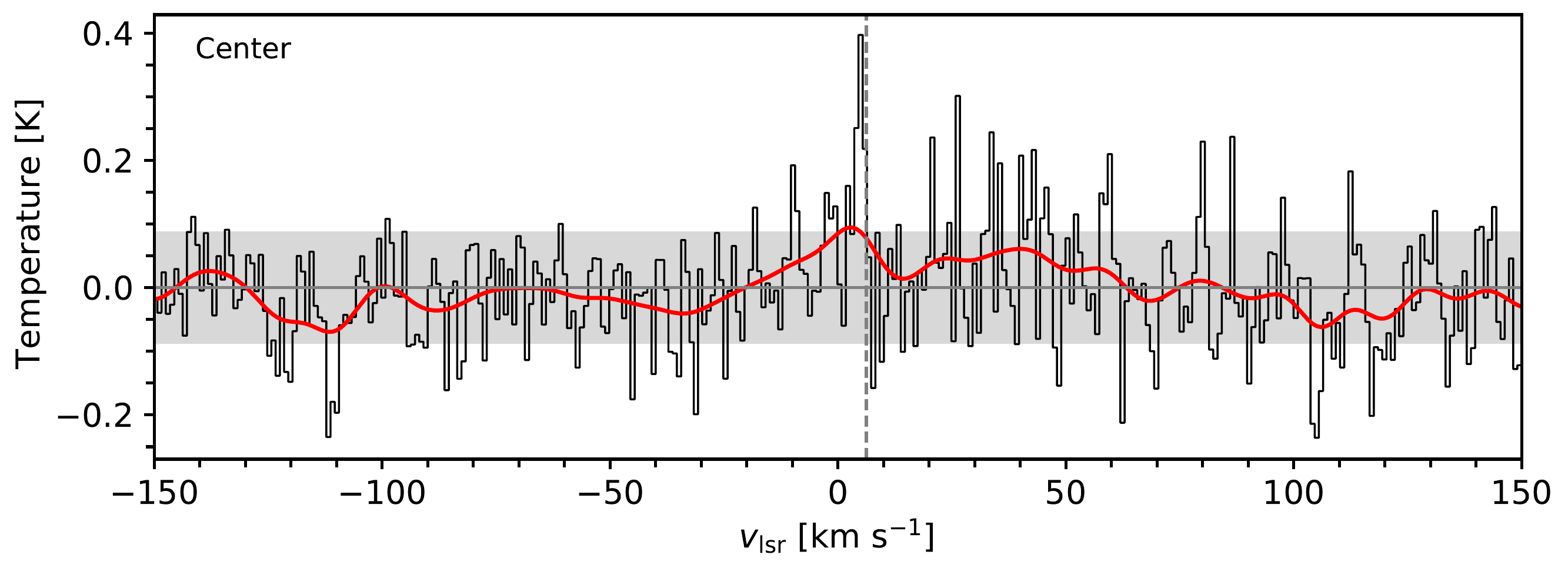}
  \includegraphics[width=0.48\textwidth]{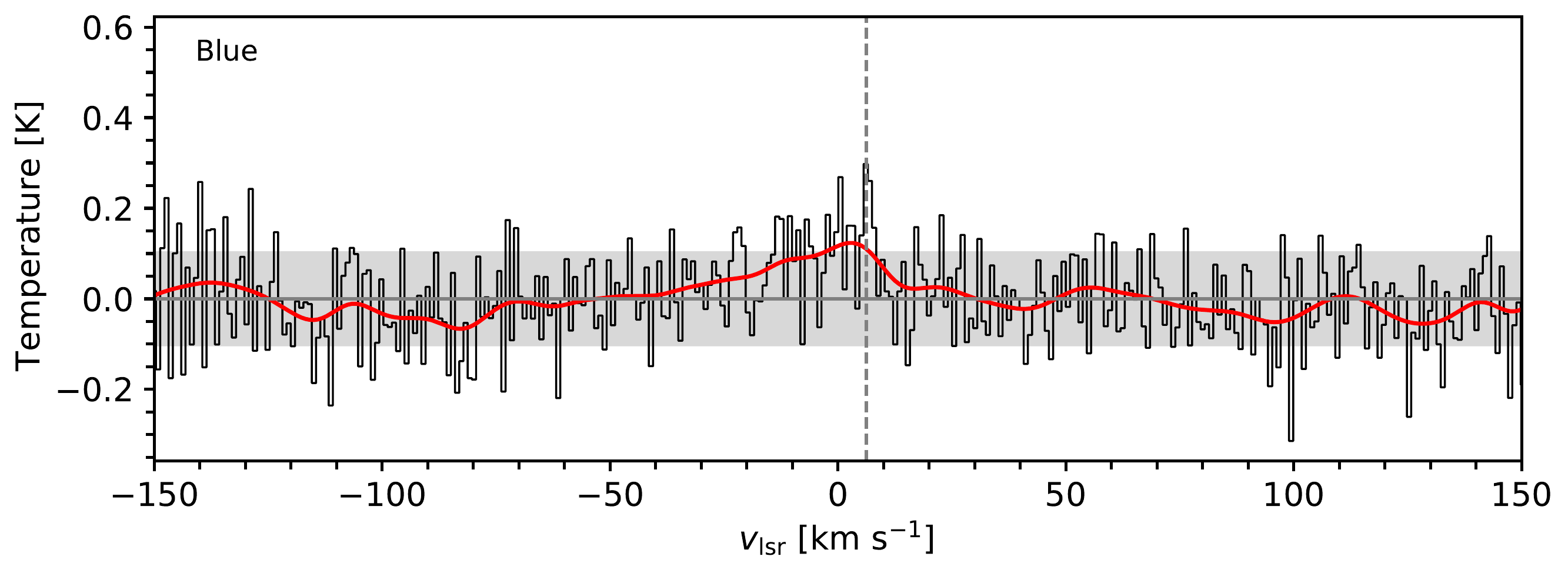}
  \caption{Same as Figure\,\ref{fig:oi}, but for the baseline-subtracted \cii\ spectra toward the ``red'', ``center'', and ``blue'' positions from top to bottom.}
  \label{fig:cii}
\end{figure}

\begin{figure*}[htbp!]
  \centering
  \includegraphics[width=0.48\textwidth]{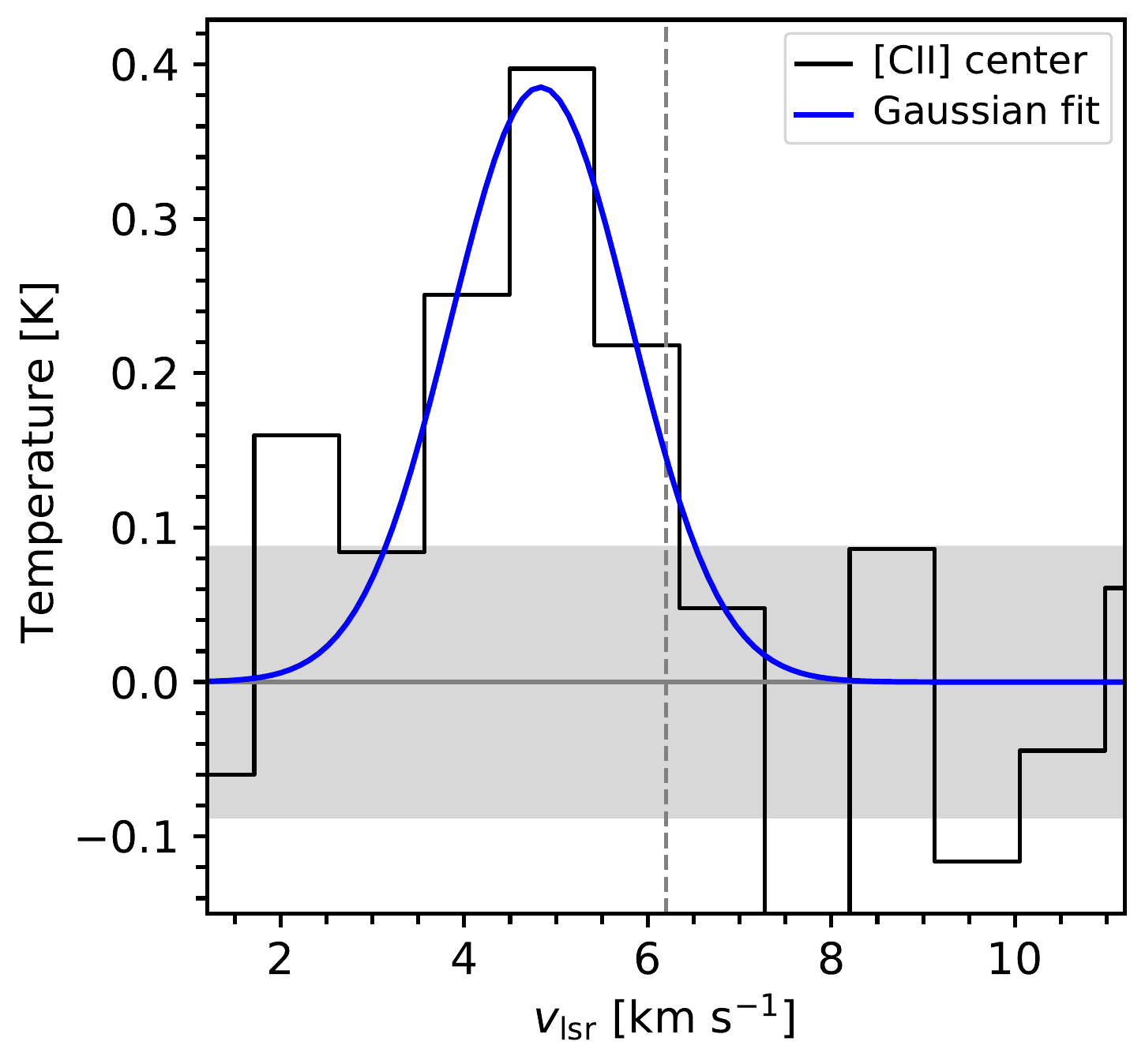}
  \includegraphics[width=0.48\textwidth]{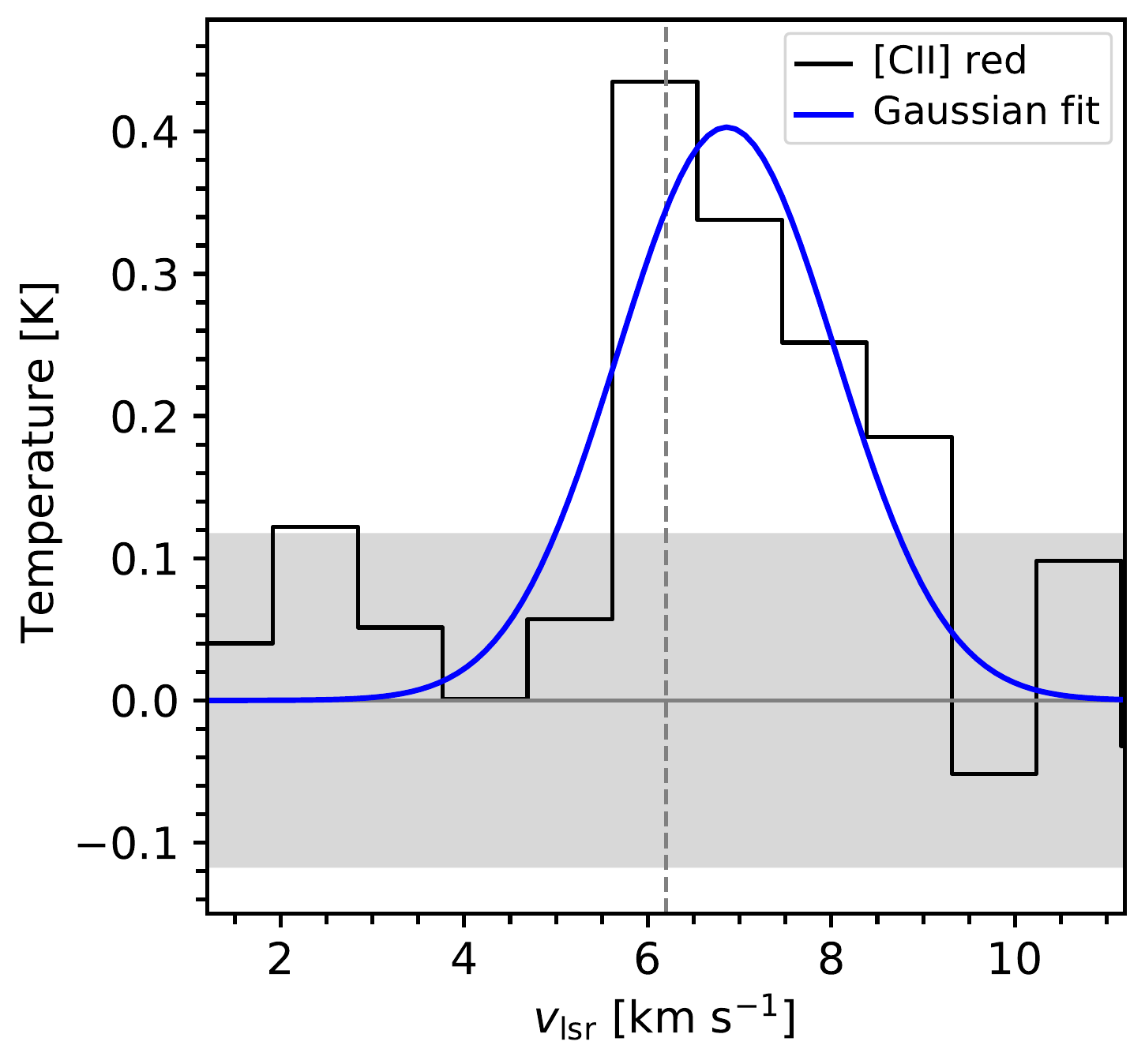}
  \caption{The zoomed-in \cii\ spectrum at the ``center'' (left) and ``red'' positions (right).  Only the low velocity emission is shown here.  The blue lines indicate the fitted Gaussian profiles.  The shaded region illustrates the $\pm1\sigma$ noise around the baseline.  See Section\,\ref{sec:result_cii} for properties of the fitted lines.}
  \label{fig:cii_lowv}
\end{figure*}

Figure\,\ref{fig:cii} shows the \cii\ spectra toward the ``red,'' ``center,'' and ``blue'' positions, where the baseline noise is 0.12 K, 0.09 K, and 0.10 K, respectively.  We used the central 80\%\ of the spectral window during baseline fitting to avoid large baseline variation toward the edge of the spectral window.  We detect a narrow feature near the source velocity in the ``center'' and ``red'' spectra (see Figure\,\ref{fig:cii_lowv} for a zoomed-in view).  These narrow features have a low signal-to-noise ratio (S/N).  By fitting a Gaussian profile, we constrain the narrow feature at 4.8$\pm$0.3\,\kms\ (``center'' position) to a width of 2.3$\pm$0.6 \kms, a peak temperature of 0.39$\pm$0.09 K, and a S/N of 4.4, while, at the ``red'' position, the narrow feature is centered at 6.9$\pm$0.3 \kms\ and is fitted with a width of 2.8$\pm$0.6 \kms, a peak temperature of 0.40$\pm$0.07 K, and a S/N of 3.4.  We also note that the feature in the ``red'' spectrum has an asymmetric line profile with the blue side of the line having a steeper profile than the red side.

\subsection{Comparison with Previous \oi\ 63 ${\it \mu}m$ and \cii\ 158 ${\it \mu}m$ Observations}
\label{sec:sofia_herschel}
Using Herschel/PACS, \citet{2016AJ....151...75G} measured line strengths of (4.2$\pm$0.1)\ee{-12} erg s$^{-1}$ cm$^{-2}$ for the \oi\ line from a spatial pixel (hereafter spaxel) centered on \source, which has a size of 9\farcs{4}$\times$9\farcs{4}.  For these SOFIA/upGREAT observations, we derived an integrated line flux at 63 \micron\ of (2.2$\pm$0.2)\ee{-12} erg s$^{-1}$ cm$^{-2}$ using a symmetric Gaussian beam ($\Omega=\theta_\text{HPBW}^2 \pi/4\text{ln}2$; $\theta_\text{HPBW}=6\farcs{3}$).  The greater line flux observed with Herschel suggests that the emission extends beyond the SOFIA beam.  If we naively assume a uniform brightness distribution and scale the SOFIA line flux to the square PACS spaxel (9\farcs{4}$\times$9\farcs{4}), the line flux becomes (4.4$\pm$0.4)\ee{-12} erg s$^{-1}$ cm$^{-2}$.  The consistency indicates that red-shifted emission is also missing in the Herschel observations (see further discussion in Section\,\ref{sec:oi_red}).

Using the Kuiper Airborne Observatory (KAO), \citet{1988ApJ...329..863C} measured an \oi\ line flux of (5.0$\pm$1.2)\ee{-12} erg s$^{-1}$ cm$^{-2}$ from a 44\arcsec\ aperture, consistent with the measurements using Herschel and SOFIA.  \citet{2016AJ....151...75G} measured a total \oi\ 63 \micron\ line flux of (7.1$\pm$0.1)\ee{-12} erg s$^{-1}$ cm$^{-2}$ over the entire 47\arcsec$\times$47\arcsec\ field of view, consistent with the KAO line flux given the larger aperture.

\citet{2016AJ....151...75G} mapped the \cii\ line flux in the 47\arcsec$\times$47\arcsec\ region around the protostar using Herschel.  From the Herschel intensity map of the \cii\ emission (Figure\,\ref{fig:footprint}, right), we can estimate the \cii\ line flux within the SOFIA beam of 14\farcs{1}, finding 1.1\ee{-13} and 1.2\ee{-13} erg s$^{-1}$ cm$^{-2}$ in the ``center'' and ``red'' positions, respectively.  The narrow feature in our SOFIA observations have a line flux of 3.5\ee{-14} and 3.1\ee{-14} erg s$^{-1}$ cm$^{-2}$ in the ``center'' and ``red'' positions, respectively, which only represents 32\%\ and 26\%\ of the total line flux detected with Herschel.  This discrepancy hints at a broad \cii\ emission undetected in our SOFIA observations or that SOFIA chopped out extended emission more than Herschel did (i.e., observations of SOFIA have more significant \cii\ emission at the chopped position than that of Herschel).  We further discuss the line width of the undetected \cii\ emission in Section\,\ref{sec:oi_origin}.

\section{Discussion}
\label{sec:discussion}
\subsection{The Missing Red-shifted \oi\ Emission}
\label{sec:oi_red}

The \oi\ emission at the ``central'' position shows only blue-shifted emission and no detectable red-shifted emission.  However, we expected to detect the emission from both lobes of outflows, resulting in a more symmetric line profile centered at the source velocity.  Given this puzzling asymmetric line profile, we consider a few scenarios that may explain our observations.  In the case of massive protostars, colder \oi\ gas in the surrounding envelope can absorb the emission from warm \oi\ originating in the outflow, resulting in absorption features around the envelope velocity, which is similar to the source velocity \citep{2014AA...562A..45K,2015AA...584A..70L}.  However, in this scenario, the emission should show a rather narrow absorption feature centered at the source velocity, which appears in our observations, and the high-velocity red-shifted \oi\ emission should have been detected (like the detected blue-shifted emission).  Thus, this scenario is unlikely.  Asymmetric outflows where the blue-shifted outflow is much stronger than the red-shifted outflow would lead to an asymmetric \oi\ line profile, such as the outflow of TMC 1A \citep{2016Natur.540..406B}.  However, the \oi\ emission observed by Herschel appears quite symmetric in both outflow lobes (Figure\,\ref{fig:footprint}, left), so this scenario is also unlikely.

\begin{figure}
  \centering
  \includegraphics[width=0.47\textwidth]{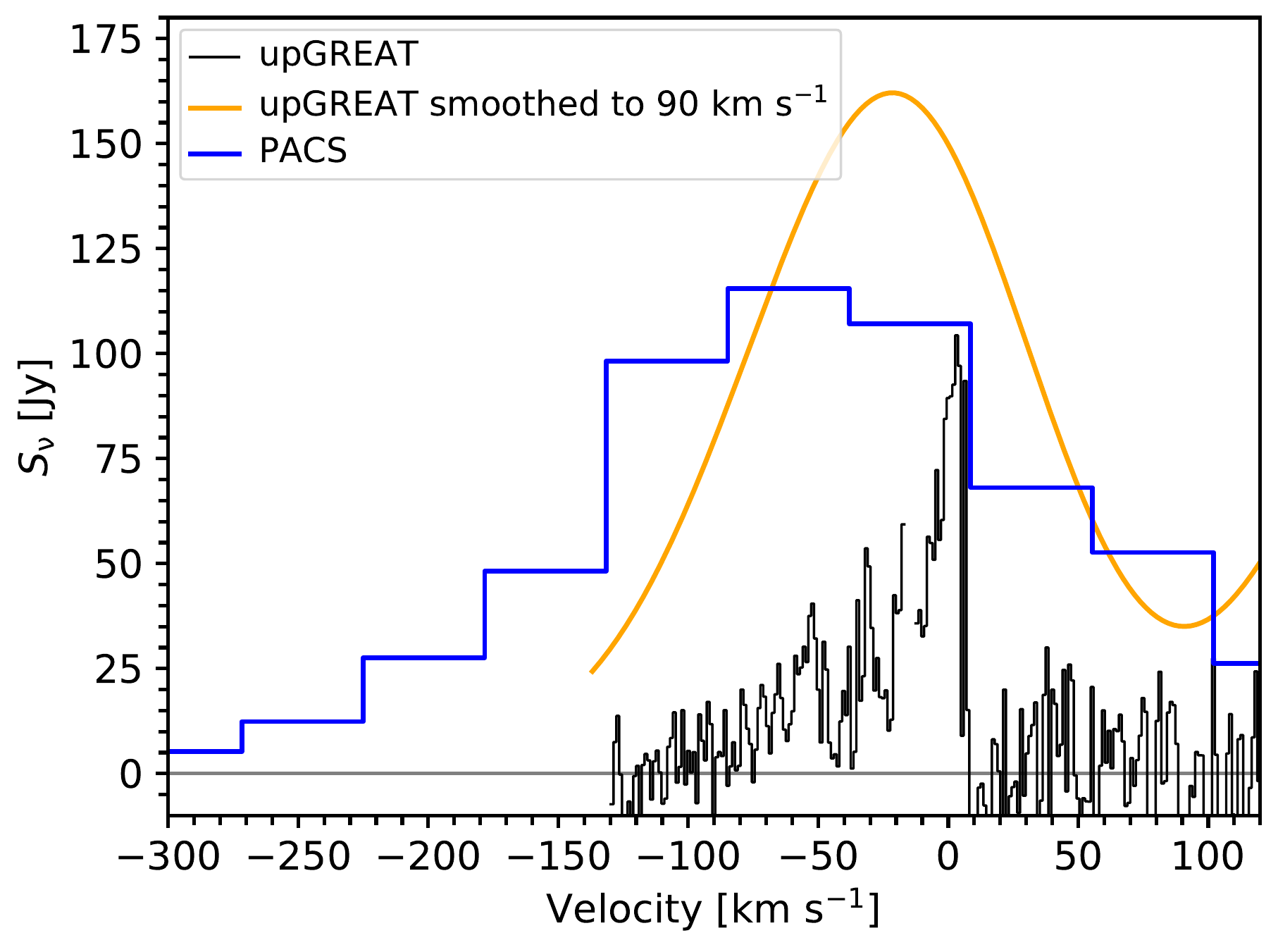}
  \caption{The \oi\ spectrum at the ``center'' position compared with the Herschel-PACS \oi\ spectrum (blue).  The upGREAT spectrum (orange) is smoothed to 90 \kms\ to match the spectral resolution of PACS.  The original upGREAT \oi\ spectrum is shown in black for comparison.}
  \label{fig:sofia_herschel}
\end{figure}

The circumbinary disk can preferentially block the red-shifted emission if most of the \oi\ emission would originate only from the inner $\sim$300 au region \citep{2019ApJ...882L...4C,2020ApJ...898...10T}.  The observed \oi\ line profile has a low-velocity ($\lesssim$20 \kms) and a high-velocity ($\sim$20--80 \kms) component (Figure\,\ref{fig:oi}; see also Section\,\ref{sec:line_profiles}).  In this scenario, both components would have a scale of $\lesssim$300 au.  Disk wind may occur at the disk scale ($r \sim 150$ au) and could produce the low-velocity component \citep{2016MNRAS.460.3472E,2020ApJ...896..126G} but not the high-velocity component, which may due to jets.  While the jets remain spatially unresolved with our SOFIA observations, the comparison of the \oi\ line profile observed by Herschel-PACS and SOFIA finds predominantly blue-shifted emission between the inner 6\farcs{2} and 9\farcs{4} regions (Figure\,\ref{fig:sofia_herschel}), suggesting that the jet extends beyond the disk scale.  The missing red-shifted emission at high-velocity indicates that the obscuration is, in fact, beyond the disk scale; thus, the circumbinary disk is unlikely to be the main obscuration mechanism.

\begin{figure}[htbp!]
  \centering
  \includegraphics[width=0.48\textwidth]{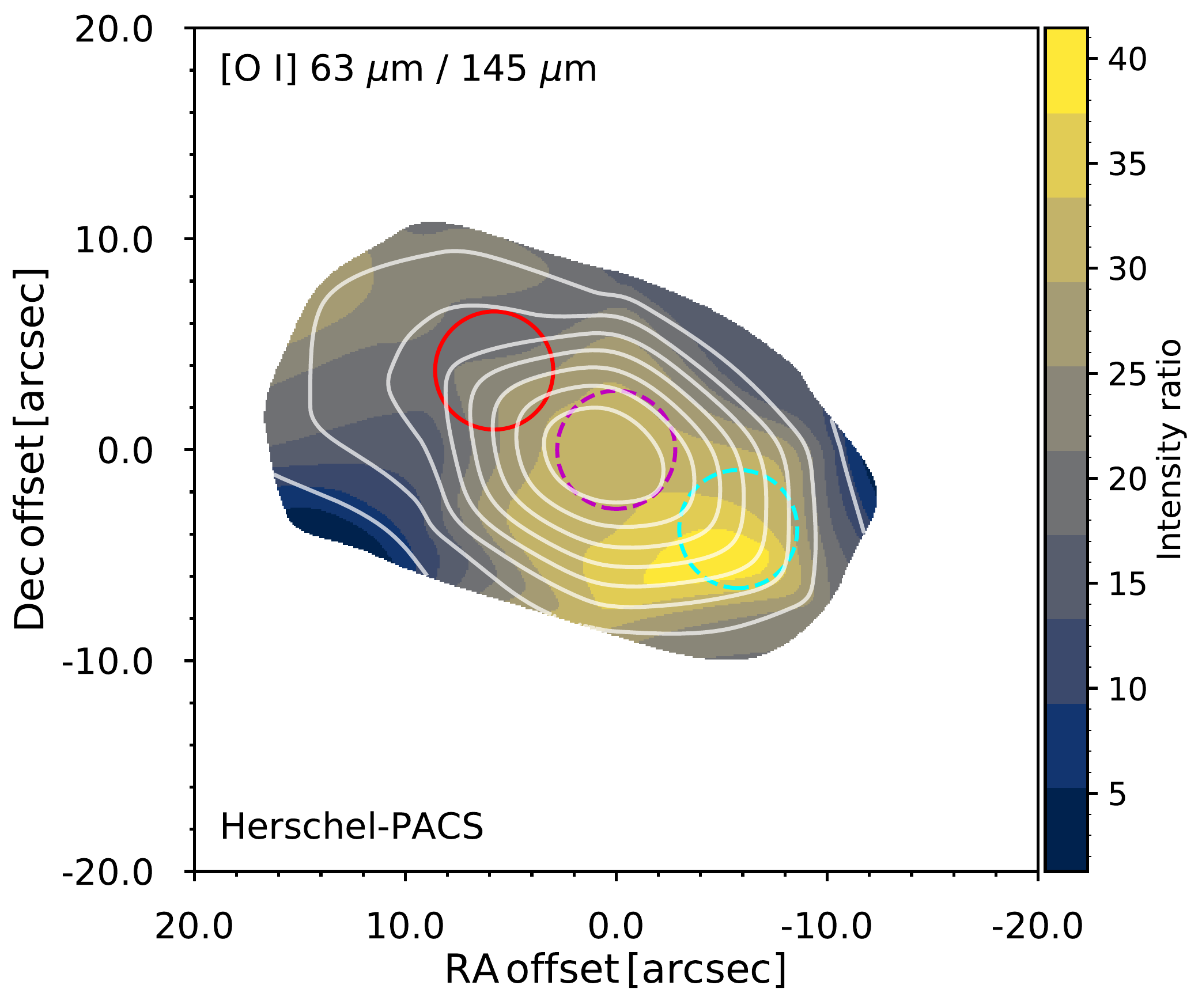}
  \caption{Intensity ratio of \oi\ 63 \micron\ to 145 \micron\ lines based on interpolating the intensity map from the Herschel observations.  The positions where the line flux is less than 20\%\ of the maximum are masked.  The white contours show the emission of \oi\ 63 \micron, while the red solid circle, magenta dashed circle, and cyan dashed circle indicate the ``red'', ``center'', and ``blue'' positions.}
  \label{fig:oi_ratio}
\end{figure}

\begin{figure}
  \centering
  \includegraphics[width=0.48\textwidth]{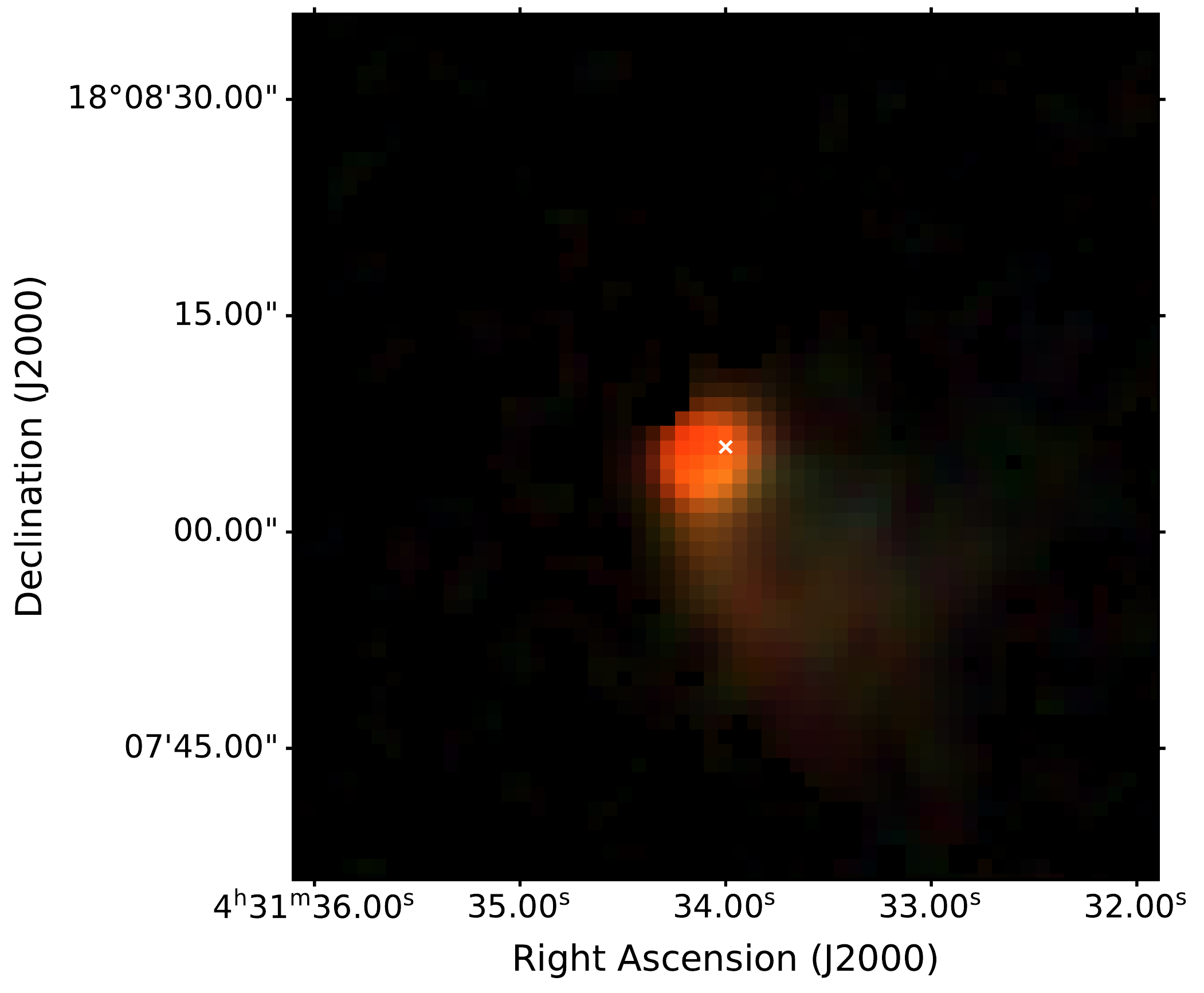}
  \caption{An RGB image of \source\ using the 2MASS $J$-, $H$-, and $Ks$-band images as blue, green, and red, respectively.  The images are combined using the method described in \citet{2004PASP..116..133L} implemented in \textsc{astropy}.}
  \label{fig:nir}
\end{figure}

Dust in the protostellar envelope can produce differential extinction in the blue- and red-shifted outflows if the outflows are inclined with respect to the plane of the sky.  The dust opacity from the envelope would impact the \oi\ 63 \micron\ line more than the \oi\ 145 \micron\ line because the dust opacity is typically proportional to $\lambda^{-\beta}$, where $\beta$ is often assumed as $\sim$1.8, making the ratio of these two lines an indicator of differential extinction.  From the Herschel observations, the ratio of the two \oi\ lines increases toward the blue-shifted outflow and decreases toward the red-shifted outflow, indicative of significant dust extinction towards the red-shifted outflow  (Figure\,\ref{fig:oi_ratio}). \citet{2015ApJ...801..121N} performed a similar analysis on other low-mass protostellar outflows but found no obvious variation of the \oi\ line ratio.  The disappearance of the red-shifted outflow in the 2MASS image of \source\ \citep{2006AJ....131.1163S} also corroborates the scenario of differential dust extinction (Figure\,\ref{fig:nir}). 

To quantitatively test the effect of differential extinction, we use the envelope model of \source\ presented in \citet{2012AA...542A...8K}, which has a density profile 
\begin{equation}
  n_\text{gas} = n_\text{in} \left(\frac{r}{r_\text{in}}\right)^{-1.8}, \text{where }r_\text{in} < r < r_\text{out}.
  \label{eq:env_model}
\end{equation}
$r_\text{in}$ and $r_\text{out}$ are the inner and outer radius of the envelope, and $n_\text{in}$ is the gas number density at $r_\text{in}$.
By fitting the SED and the radial brightness profiles at 450\,\micron\ and 850\,\micron, \citet{2012AA...542A...8K} constrained these parameters to $n_\text{in} \sim$ 6.9\ee{8}\,\cc, $r_\text{in} \sim$ 28.9 au, and $r_\text{out} \sim$ 14,000 au.  To model the optical depth toward the blue- and red-shifted outflows, we assume triangular outflow cavities and adopt geometric parameters from the literature.  

For example, we adopt an outflow half-opening angle ($\theta_\text{cav}$) of 22$^{\circ}$ \citep{2009ApJ...698..184W} and an inclination angle ($\theta_\text{incl}$) of 30$^{\circ}$ with respect to the plane of the sky \citep{2014ApJ...796...70C}.  Then, to derive the column density ($N_\text{gas}$) at any impact parameter ($b$), we can simply integrate the density profile (Equation\,\ref{eq:env_model}) from either the inner edge of the envelope, $\sqrt{r_\text{in}^2-b^2}$, or the outflow cavity wall, $b\,\text{tan}(\theta_\text{cav}+\theta_\text{incl})$, to the outer edge of the envelope, $\sqrt{r_\text{out}^2-b^2}$.  The $\theta_\text{cav}$ is positive for the blue-shifted outflow cavity and negative for the red-shifted outflow cavity.  With an estimate of the column density in hand, we can estimate the dust optical depth from
\begin{equation}
  \tau_\text{dust} = \kappa_\text{63\,$\mu$m} N_\text{gas} \mu m_\text{H} 0.01,
\end{equation}
where $\kappa_\text{63\,$\mu$m} =$ 2.1\ee{2} cm$^{-2}$ g$^{-1}$ is the dust opacity at 63\,\micron\ taken from Table 1, column 5 in \citet{1994AA...291..943O}, $\mu = 2.8$ is the mean molecular weight \citep{2008AA...487..993K}, $m_\text{H}$ is the mass of a hydrogen atom, and 0.01 is the dust-to-gas mass ratio.  

\begin{figure}[htbp!]
  \centering
  \includegraphics[width=0.48\textwidth]{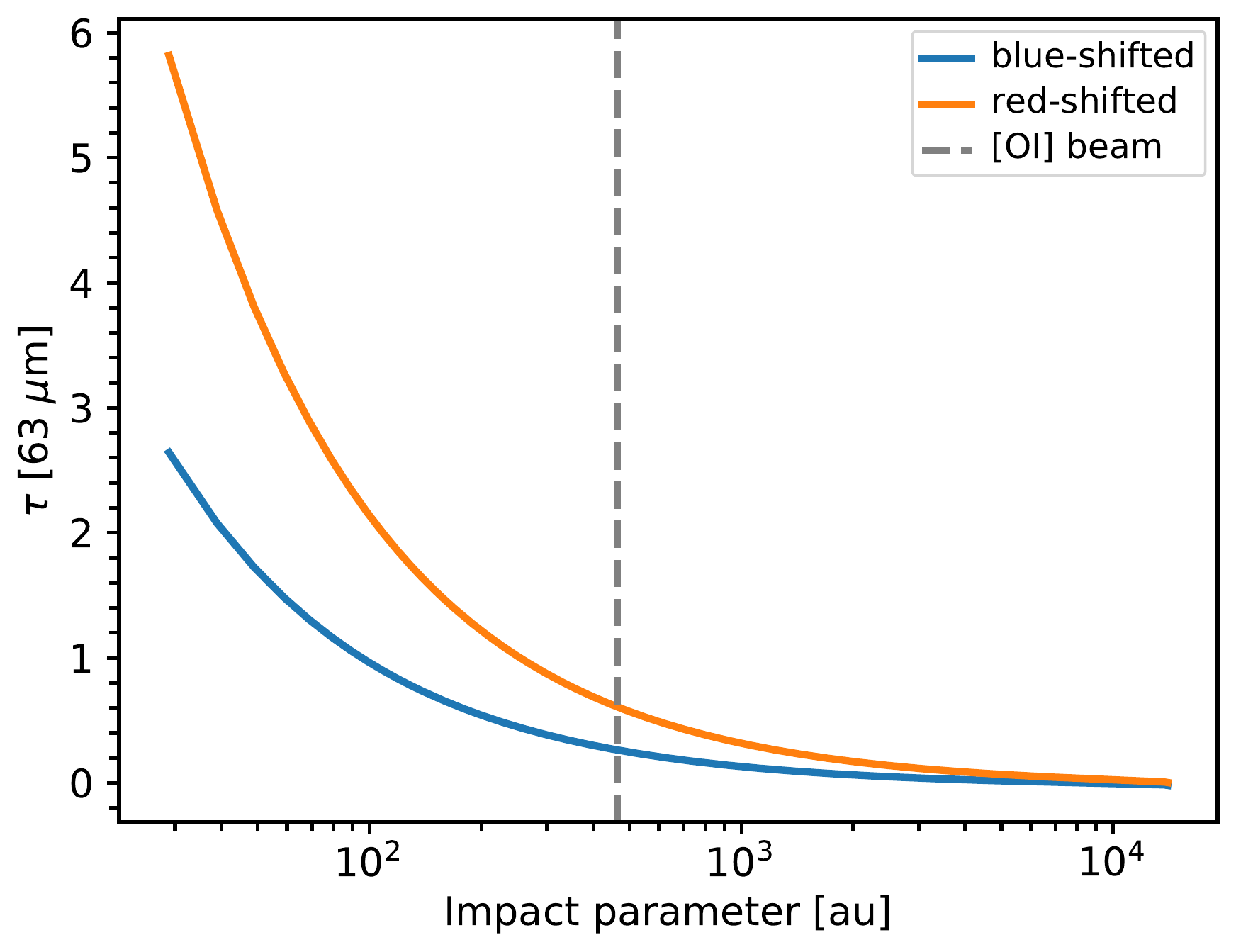}
  \caption{Dust optical depth as a function of impact parameters in the blue- and red-shifted outflows.  The gray dashed line indicates the beam radius of the SOFIA observations, 3\farcs{15}, at a distance of 147.3 pc.}
  \label{fig:optical_depth}
\end{figure}

\begin{figure}[htbp!]
  \centering
  \includegraphics[width=0.48\textwidth]{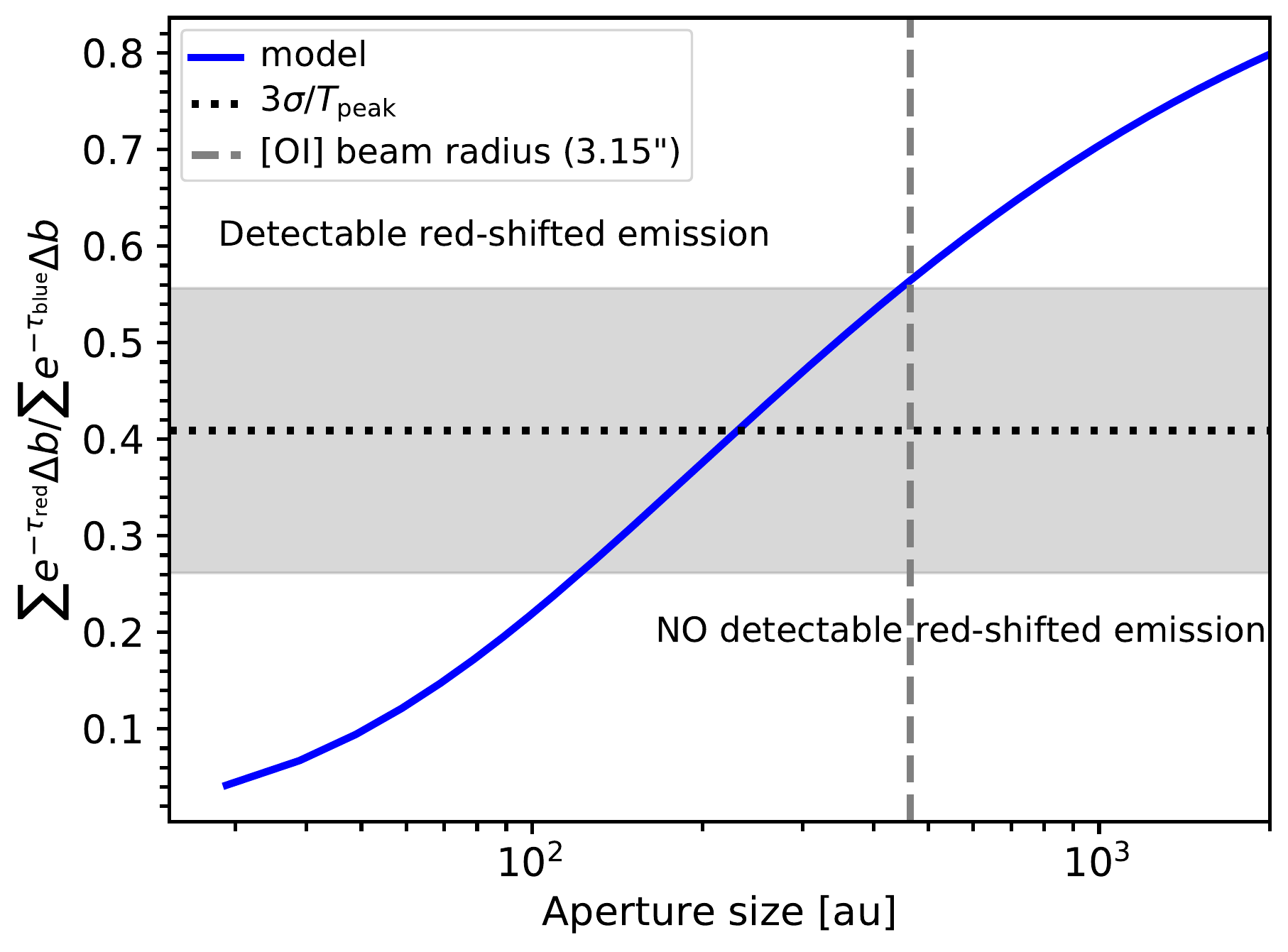}
  \caption{Ratio of the integrated optical depth correction factors in the red- and blue-shifted outflows as a function of aperture size.  The gray dashed line indicates the beam radius at 63 \micron.  The dotted line indicates the ratio of the $3\sigma$ limit on the \oi\ emission to the peak intensity of the emission at the ``center'' position, while the shaded region indicates the uncertainty of the ratio.}
  \label{fig:optical_depth_ratio}
\end{figure}

Because the outflow is inclined, the integrated dust optical depth toward the red-shifted outflow is greater than that toward the blue-shifted outflow (Figure\,\ref{fig:optical_depth}).  Assuming both outflows have the same intrinsic \oi\ line flux, the observed flux ratio between two lobes can be approximated by the ratio of $\Sigma e^{-\tau}\Delta b$, where $\tau$ is the dust optical depth in each outflow lobe as presented in Figure\,\ref{fig:optical_depth}.  The results are given in Figure\,\ref{fig:optical_depth_ratio}, which shows that the dust extinction from the envelope can significantly reduce the flux ratio to $\sim$0.55$\pm$0.15 with a SOFIA beam size of 6\farcs{3}.  The observed flux ratio upper limit agrees with the modeled ratio within the uncertainty, but is lower than the modeled ratio.  If the \oi\ emission originates from shocked gas, the \oi\ line intensity would be higher toward the protostar, resulting in a lower flux ratio in the model that assumes a uniform brightness.  With a resolution of 9\farcs{4}, the Herschel/PACS observations (Figure\,\ref{fig:footprint}) cannot resolve the inner $\sim6\arcsec$ where the red-shifted outflow is mostly extincted ($\tau > 1$ in Figure\,\ref{fig:optical_depth}).  Beyond the inner 9\farcs{4} region, the dust extinction has negligible impact on the \oi\ emission, resulting in a rather symmetric distribution of \oi\ emission in Figure\,\ref{fig:footprint}.  
Thus, extinction due to the dusty envelope surrounding \source\ is the most likely reason for the absence of red-shifted emission in the resolved \oi\ line, although the circumbinary disk may block some low-velocity red-shifted \oi\ emission.  The asymmetric line profiles of CO\,\jj{16}{15} and \water\ $2_{02}\rightarrow1_{11}$, which are discussed in Section\,\ref{sec:line_profiles}, also corroborate with the extinction due to the envelope.  As the wavelength increases, from the CO line to the \water\ line, the asymmetry becomes less obvious (Figure\,\ref{fig:optical_depth_ratio}).

\begin{figure}
  \centering
  \includegraphics[width=0.48\textwidth]{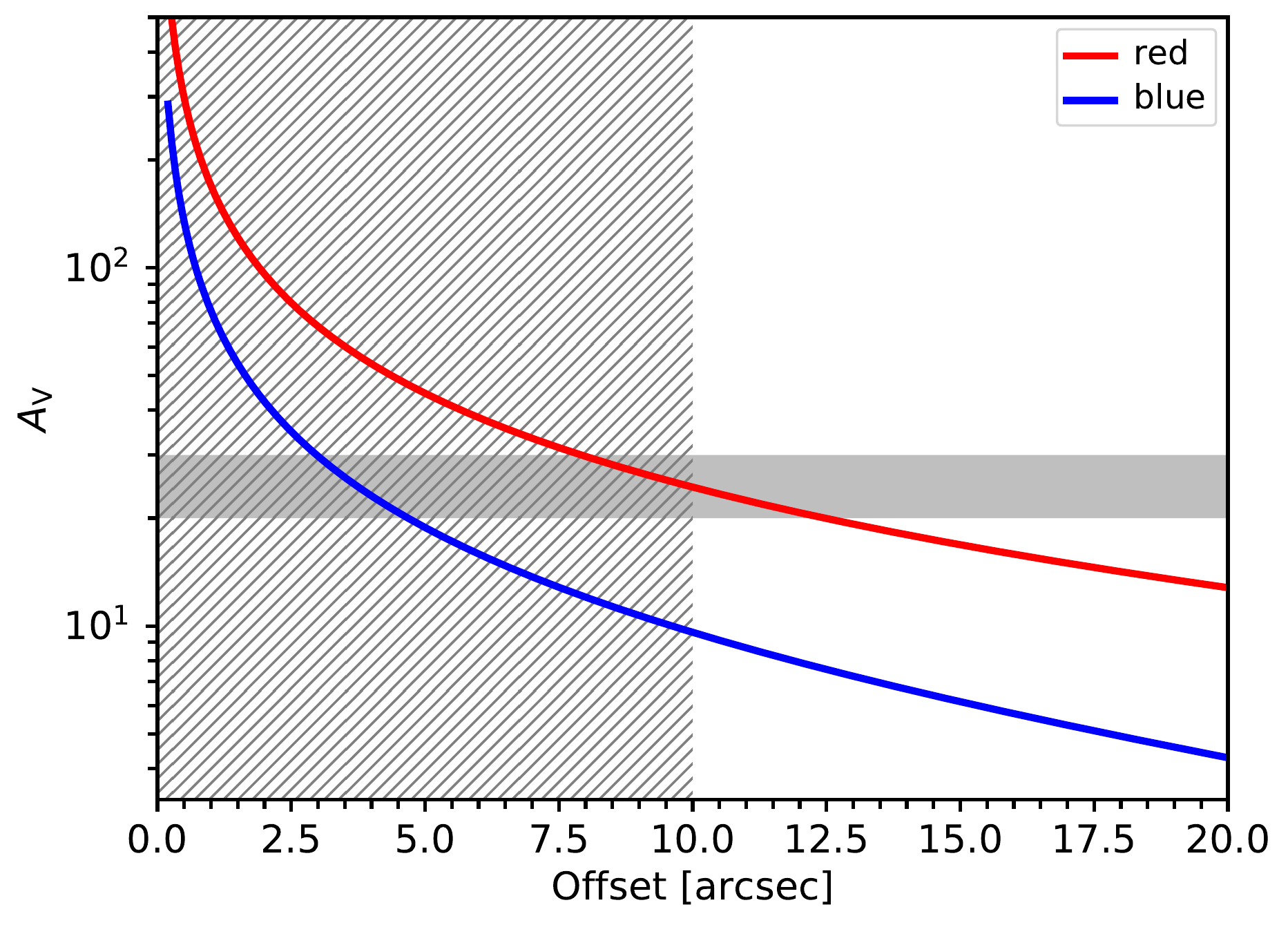}
  \caption{Visual extinction ($A_{\rm V}$) as a function of radius for the envelope model.  The red and blue lines show the $A_{\rm V}$ along the red- and blue-shifted outflows, respectively.  The hatched region highlights the area where \citet{2009ApJ...694..582H} found no signature of jets toward the red-shifted outflow and the gray area indicates their estimated $A_{\rm V}$ from the faint jet detected at $>$10\arcsec\ from the protostar.}
  \label{fig:av}
\end{figure}

We further test the extinction from the envelope model using observations of \feii\ jets in \source.  \citet{2009ApJ...694..582H} detected a faint \feii\ jet in the red-shifted outflow, appearing only at separations $>$10\arcsec\ from the protostar.  They also measured a visual extinction, $A_{\rm V}$, of 20--30 mag for this faint jet.  We estimate the $A_{\rm V}$ using the dust opacity at 0.55 \micron\ (2.1\ee{4} cm$^2$ g$^{-1}$) and taking $A_{\rm V} = 2.5\text{log}_{10}(e^{\tau_{\lambda}}) = 1.086 \tau_{\rm V}$. Figure\,\ref{fig:av} shows the $A_{\rm V}$ along the red- and blue-shifted outflows.  At an offset $>$10\arcsec, the $A_{\rm V}$ toward the red-shifted outflows is consistent with the $A_{\rm V}$ measured from the \feii\ jet, supporting the scenario of envelope extinction for the \oi\ line.

\subsection{Shocks as the Origin of the \oi, CO, and \water\ Emission}
\label{sec:line_profiles}

\begin{figure*}[htbp!]
  \centering
  \includegraphics[width=\textwidth]{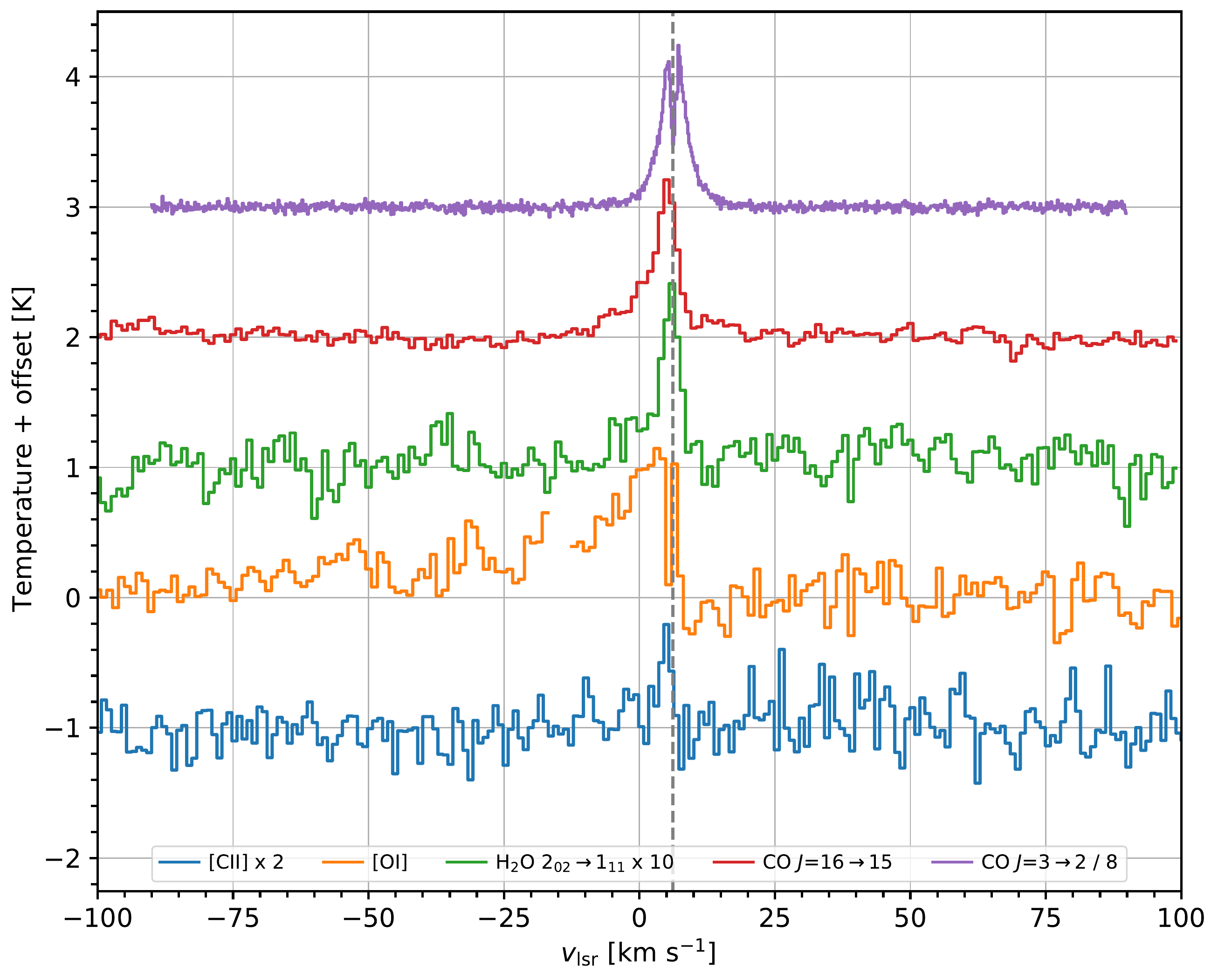}
  \caption{Spectra of CO\,\jj{3}{2}, CO\,\jj{16}{15}, \water\ $2_{02}\rightarrow1_{11}$, \oi\ 63 \micron, and \cii\ 158 \micron\ lines toward the ``center'' position (from top to bottom).  Each spectrum is offset by 1 K for better visualization.  The dashed line indicates the source velocity of 6.2 \kms.}
  \label{fig:all_comparison}
\end{figure*}

The broad line width and significant velocity offset of the \oi\ line profile suggests an outflow origin.  From observations of far-infrared emission in outflows, a composite scenario emerges from high-$J$ CO and \water\ lines \citep[e.g.,][]{2012AA...542A...8K,2017AA...605A..93K,2017AA...600A..99M,2018ApJS..235...30K,2018ApJ...860..174Y}.  In this scenario, the outflow-envelope interaction leads to shock-excited emission of high-$J$ CO, \water, and \oi\ lines.  So-called ``cavity'' shocks at the envelope-outflow interface can produce broad (typical FWHM of $>15$ \kms) emission centered at the source velocity, while ``spot'' shocks occurring within the outflow (typical FWHM of $=5-15$ \kms) result in emission that is offset from the source velocity, and often blue-shifted.  Magneto-hydrodynamical disk winds can also produce similar features as the cavity shocks \citep{2016AA...585A..74Y}.  Some sources show a component similar to that of spot shocks but with significant velocity offset, and are often called Extremely High Velocity (EHV) gas or ``bullets''.  Both components are thought to originate from spot shocks, likely tracing dissociative J-shocks \citep{2014AA...572A..21M,2021AA...648A..24V}.  
Colder molecular gas entrained by outflows along the edges of the cavity mostly dominates the emission of low-$J$ CO lines, such as CO\,\jj{3}{2}, showing a narrow (FWHM $<5$ \kms) line profile centered on the source velocity \citep[][Figure\,\ref{fig:all_comparison}]{2013AA...556A..89Y}.

\begin{figure*}[htbp!]
  \centering
  \includegraphics[width=\textwidth]{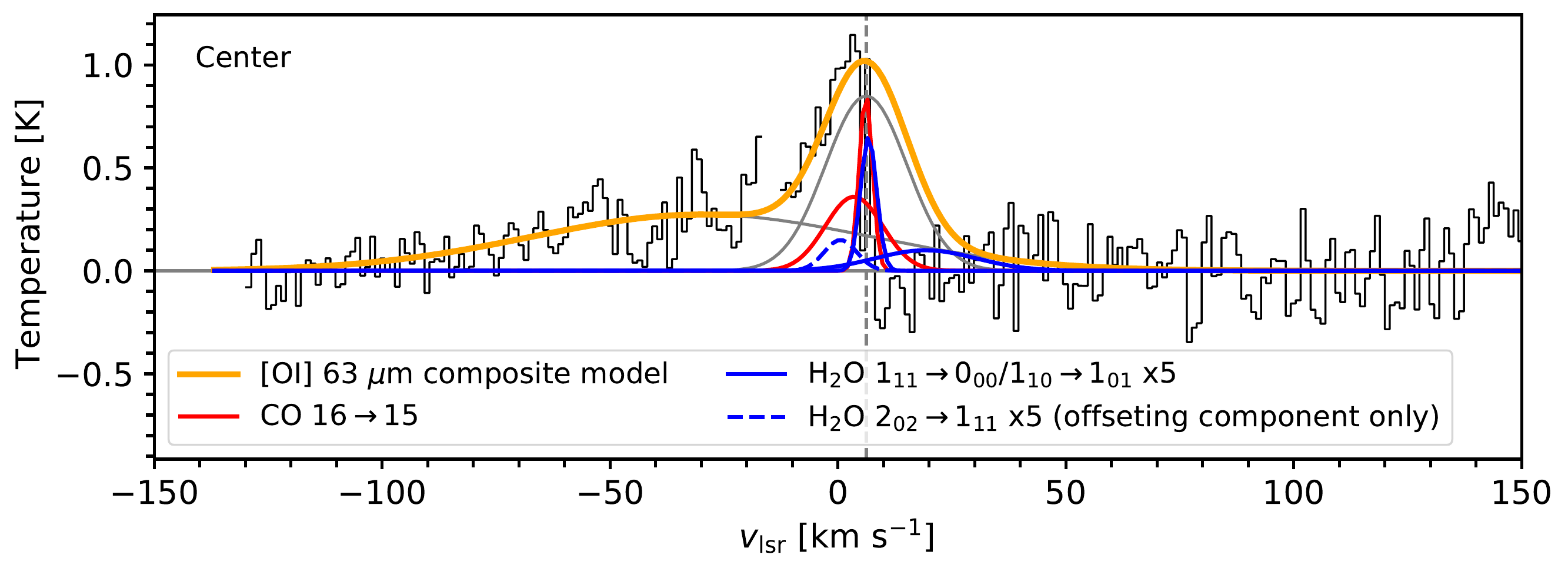}
  \caption{The ``central'' \oi\ spectrum (black) along with the best-fitting double Gaussian model.  The gray lines indicate each Gaussian component, while the orange line shows the composite model.  The FWHMs of the two components are 21.0$\pm$4.9 \kms\ and 87.5$\pm$32.3 \kms, respectively.  The red lines show the decoupled components of the CO\,\jj{16}{15} line from \citet{2017AA...605A..93K} with a narrow component with a width of 3.6 \kms\ at 6.1 \kms\ and a relatively broad component with a width of 14.9 \kms\ at 3.4 \kms.  The blue lines show the decoupled components of the multiple \water\ lines.  \citet{2014AA...572A..21M} derived a narrow component at 6.7 \kms\ with a FWHM of 4.2 \kms\ and a red-shifted broad component at 19.2 \kms\ with a FWHM of 26.1 \kms\ from the \water\ $1_{11}\rightarrow0_{00}$ and $1_{10}\rightarrow1_{01}$ lines (solid blue lines).  The blue-shifted component centers on 0.54 \kms\ with a FWHM of 8.4 \kms, fitted from the \water\ $2_{02}\rightarrow1_{11}$ line using \textsc{class} (blue dashed line).  The intensity of the \water\ components are multiplied by five for a better visualization.  The vertical dashed line indicates the source velocity at 6.2 \kms.}
  \label{fig:oi_fitting}
\end{figure*}

The \oi\ 63 \micron\ line profile in the ``center'' position seems to have two components as well, including a component centered at the source velocity with its red-shifted part blocked by the dusty envelope (see discussion in Section\,\ref{sec:oi_red}) and a broad component at high blue-shifted velocity.  Assuming a two-component model, we fitted two Gaussian profiles with one component fixed at the source velocity, 6.2 \kms, and constrained another broad component at $-30.0$ \kms\ (Figure\,\ref{fig:oi_fitting}).  Table\,\ref{tbl:fitting} lists the fitted parameters of the two Gaussian profiles.  This composite model successfully reproduces the observed line profile, suggesting two distinct components of \oi\ gas; however, the fitted parameters of the extremely broad component have $\sim$35--60\%\ uncertainties.

\citet{2017AA...605A..93K} decomposed the Herschel/HIFI CO\,\jj{16}{15} emission in \source\ into two components tracing different physical components. The component that is centered around the source velocity has a narrow line width of 3.6 \kms, and traces the entrained gas. Another component at 3.6 \kms\ has a width of 14.9 \kms\ and traces the cavity shocks (Figure\,\ref{fig:oi_fitting}).  In comparison, the \oi\ emission has no narrow component in emission but does show a narrow absorption near the source velocity (at $\sim$5.4 \kms).  This narrow absorption may come from the colder \oi\ gas in the quiescent or the infalling envelope.  The broad component in CO\,\jj{16}{15} has a similar line profile to the \oi\ component centered at the source velocity.  The centroid of the CO broad component is 3.5 \kms, which is consistent with the \oi\ component centered at the source velocity given the uncertainty in the \oi\ spectrum, suggesting that both lines trace the component.

Four \water\ lines were detected by Herschel \citep{2012AA...542A...8K,2014AA...572A..21M}.  The $2_{02}\rightarrow1_{11}$ and $2_{11}\rightarrow2_{02}$ lines show a narrow component (FWHM $=4.3\pm0.8$ \kms) centered at the source velocity with a blue-shifted line wing (the \water\ $2_{02}\rightarrow1_{11}$ emission is shown in Figure\,\ref{fig:all_comparison}).  However, the other two \water\ lines, $1_{11}\rightarrow0_{00}$ and $1_{10}\rightarrow1_{01}$, show a broad (FWHM $=26.1\pm6.3$ \kms) red-shifted component at $19.2\pm3.0$ \kms\ but no narrow component.  Figure\,\ref{fig:oi_fitting} also shows the decoupled components derived from \water\ lines \citep{2014AA...572A..21M} and the blue-shifted component from a double-Gaussian fit to the \water\ $2_{02}\rightarrow1_{11}$ line (Figure\,\ref{fig:oi_fitting}).  At blue-shifted velocities, the \water\ lines in \source\ support a similar scenario as the CO\,\jj{16}{15} line, where the emission comes from entrained gas and cavity shocks traced by the blue-shifted line wing.  At red-shifted velocities, the \water\ lines show a velocity component at $19.2\pm3.0$ \kms\ with a width of $26.1\pm6.3$ \kms, suggesting an origin in cavity shocks \citep{2014AA...572A..21M}.  The reason why the four \water\ lines exhibit different kinematics remains unclear.  A slightly blue-shifted narrow absorption also appears in the $1_{11}\rightarrow0_{00}$ \water\ line, indicative of absorption by a colder envelope \citep{2014AA...572A..21M}.

Among the spectra of high-$J$ CO, \water, and \oi, only the \oi\ emission has the extremely broad component with a width of $\sim$90 \kms.  Combining the insights learned from the \oi\ 63 \micron\ and CO\,\jj{16}{15} lines, a revised picture of the \source\ outflow emerges:  The narrow (FWHM = 3.6 \kms) component traces entrained material that appears in emission for CO\,\jj{16}{15} and \water\ \citep{2014AA...572A..21M}.  The quiescent envelope appears as narrow absorption seen in both the \water\ $1_{11}\rightarrow0_{00}$ line and the \oi\ 63 \micron\ line.  The broad component (FWHM $\sim$20 \kms) comes from the cavity shocks where the outflow drives shocks into the cavity walls.  The extremely broad component (FWHM $\sim$90 \kms) that is only seen in \oi\ traces J-type spot shocks in the outflow cavity or the jet.  The line profiles of all three species, \oi, high-$J$ CO, and \water, show asymmetries that agree with the extinction due to the envelope (see Section\,\ref{sec:oi_red}).
 
\citet{2016AJ....151...75G} detected \oi\ emission at 63 and 145 \micron\ using Herschel/PACS, further constraining the excitation condition of \oi\ in \source.  The ratio of \oi\ 63 \micron/145 \micron\ is $32.0\pm4.4$ for the central spaxel and $25.4\pm2.5$ for the 1D spectrum extracted with a 24\farcs{8} aperture, which best matches the SPIRE spectrum \citep[][see also Figure\,\ref{fig:oi_ratio} for the map of ratios]{2018ApJ...860..174Y}. \citet{2015ApJ...801..121N} modeled the ratio of the \oi\ 63 \micron\ to 145 \micron\ lines, which increases with the density of collision partners, suggesting a H$_2$ density of $\sim$10$^{5.5}$\,\cc\ for the \oi\ outflow.

The study of the R1 position in the outflow of NGC 1333 IRAS 4A is a notable comparison to \source\ \citep{2017AA...601L...4K}.  In R1, the \oi\ 63 \micron\ emission observed by SOFIA has a similar line profile as that of \water\ and high-$J$ CO lines, which are tracers of shocks, suggesting that this emission lines mostly come from the outflows.  In contrast, the \water\ line in \source\ has a different line profile compared to the \oi\ spectrum, suggesting that the \oi\ emission uniquely trace an atomic outflow and/or jet.  

\begin{table}
  \centering
  \caption{The fitted Gaussian components in the ``central'' \oi\ emission \label{tbl:fitting}}
  \begin{tabular}{ccc}
    \toprule
    Velocity centroid & Integrated flux & Width \\
    (\kms) & (K \kms) & (\kms) \\
    \midrule
    6.2$^{a}$      & $18.9\pm6.3$  & $21.0\pm4.9$ \\
    $-30.0\pm19.6$ & $25.4\pm11.3$ & $87.5\pm32.3$ \\
    \bottomrule
    \multicolumn{3}{l}{$^{a}$The centroid is fixed to the source velocity.}
  \end{tabular}
\end{table}

\subsection{The Spatial Extent of Shocks and its Connection to UV Radiation}
\label{sec:oi_origin} 

Both shocked gas in outflows and photodissociation regions (PDRs) can contribute to the \oi\ emission \citep{1985Icar...61...36H,1985ApJ...291..722T}. Although our SOFIA observations did not detect significant \oi\ emission at ``red'' and ``blue'' outflow positions, \citet{2016AJ....151...75G} detected extended \oi\ emission with Herschel/PACS on $\sim$20\arcsec\ scales (Figure\,\ref{fig:footprint}).  Interpolating from the Herschel \oi\ emission map, we estimate \oi\ velocity-unresolved fluxes of 7.7\ee{-13} and 7.6\ee{-13} erg s$^{-1}$ cm$^{-2}$ in the ``red'' and ``blue'' positions, respectively.  If the \oi\ flux observed by Herschel comes from a single line, our SOFIA observations have a better sensitivity to detect a narrow (FWHM $\lesssim5$ \kms) line from PDRs compared to a broad line due to shocks.  The non-detection of narrow \oi\ emission in all three pointings allows us to constrain the maximum contribution to the \oi\ emission from PDRs by comparing the non-detection to the interpolated \oi\ flux from Herschel/PACS.  \cii\ emission is a common tracer of PDRs \citep{1999RvMP...71..173H,1999ApJ...527..795K}.  If we take the mean FWHM of the narrow \cii\ lines, 2.5 \kms, as the characteristic line width of gas in PDRs, the upper limit of the \oi\ line flux is then 2.4\ee{-14} erg s$^{-1}$ cm$^{-2}$ at both outflow positions and 3.7\ee{-14} erg s$^{-1}$ cm$^{-2}$ at the central position.  Comparing to the Herschel \oi\ line fluxes estimated within the ``central'' (from Section\,\ref{sec:sofia_herschel}), ``red'', and ``blue'' positions, the upper limit of the PDR contribution to the \oi\ flux is 3.1\%\ in ``red'' and ``blue'' positions, while the contribution to the \oi\ flux is 2.6\%\ in the ``central'' position, suggesting that the shocked gas (i.e., broad emission) dominates the \oi\ flux.  Assuming that a single Gaussian line profile can describe all the non-PDR \oi\ emission, \oi\ emission at ``red'' and ``blue'' positions should have FWHM greater than $\sim$80 \kms had it been detected, which is similar to the line width of the extremely broad \oi\ component at the ``central'' position.

The narrow line width of the observed \cii\ emission indicates a PDR origin.  Extracting the line flux from the Herschel intensity map (Figure\,\ref{fig:footprint}, right) using the SOFIA beam (14\farcs{1} at the \cii\ line), the detected narrow feature makes up 30\%\ and 26\%\ of the total Herschel \cii\ flux at the ``center'' and ``red'' positions, respectively.  If the \cii\ emission from the large-scale cloud is removed by beam-chopping (see Section\,\ref{sec:obs}), the detected \cii\ line flux then traces the emission from local PDRs.  The PDRs contribute $\sim$30\%\ of the total \cii\ flux but only represents $\lesssim3\%$ of the \oi\ flux, suggesting that \oi\ emission is indeed a good tracer of shocked gas and the contribution from PDRs is negligible.  Furthermore, the spatial extent of the \oi\ emission coincides with the blue-shifted \feii\ emission that unambiguously traces dissociative shocks in the jets \citep{2009ApJ...694..654P}.  If the discrepant \cii\ emission between observations of SOFIA and Herschel is instead due to excessive baseline subtraction during data reduction, we can assume a Gaussian profile for the undetected \cii\ emission and estimate the lower limit of its line width given the spectral noise.  We estimate that the \cii\ flux missing in the SOFIA observations is 7.7\ee{-14} and 8.9\ee{-14} erg s$^{-1}$ cm$^{-2}$, resulting in a lower limit of 23.3 and 20.2 \kms\ for the line width at the ``center'' and ``red'' positions, respectively.  Comparing to the lower limit of the \oi\ line width, the undetected \cii\ emission could be narrower than the undetected \oi\ emission.  

With the PDR contribution constrained toward the protostar, we can further use the line ratio between \cii\ and \oi\ to infer the UV radiation incident on the PDR.  \citet{2018ApJS..235...30K} calculated the ratio between the \cii\ 158 \micron\ and the \oi\ 63 \micron\ lines as a function of $G_{0}$ from \eten{1}--\eten{6.5} and densities from \eten{3}--\eten{7} \cc.  For the PDR contribution in the ``center'' position, we measure a \cii\ line flux of 3.3\ee{-14} erg s$^{-1}$ cm$^{-2}$ in a 14\farcs{1} beam, and an upper limit of the \oi\ line flux of 3.7\ee{-14} erg s$^{-1}$ cm$^{-2}$ in a 6\farcs{3} beam.  Thus, if we scale the \oi\ upper limit to the \cii\ beam size, we have a lower limit of 0.17 on the ratio of \cii\ and \oi.  The corresponding \cii\ integrated line intensity is 9.0\ee{-6} erg s$^{-1}$ cm$^{-2}$ sr$^{-1}$ and, combining with the \cii\ intensity, we estimate that the incident UV radiation, $G_{0}$, is at most $\sim$10 for a density of $\sim$\eten{5} \cc, which is similar to the density estimated from the ratio of \oi\ 63 \micron\ and 145 \micron\ lines (see Section\,\ref{sec:line_profiles}; \citealt{2015ApJ...801..121N}). 

A strength of $\sim$10 for the UV radiation is consistent with prior estimates for low-mass protostars, which assumed that 90\% of unresolved \oi\ emission originates from shocks and 10\% from PDRs \citep{2018ApJS..235...30K}. The velocity-resolved line profiles from SOFIA-upGREAT verify such assumptions by identifying the narrow and broad components in \oi\ and \cii\ lines.

\subsection{Chemical Abundance}
\label{sec:abundance}

Besides being a good tracer of energetic outflows and jets, atomic oxygen is a major carrier of volatile oxygen in protostellar shocks.  The Herschel Key Program, Water In Star-forming regions with Herschel (WISH), demonstrated that CO, \water, and atomic O represent most of the volatile oxygen budget in protostellar outflows and shocks, but up to 50\%\ of the total oxygen abundance remains unaccounted for \citep[][and references therein]{2021AA...648A..24V}.  Velocity-resolved spectra of \oi, CO\,\jj{16}{15}, and \water\ give us a unique opportunity to identify the emission from shocks as well as to quantify the column density of each oxygen carrier in shocks.  The two components of the \oi\ emission (Table\,\ref{tbl:fitting}) represent two types of shocks.  The broad component (FWHM $\sim15-25$ \kms) seen in \oi, CO\,\jj{16}{15}, and \water\ trace cavity shocks, while the extremely broad component only seen in \oi\ probes the shocked gas in jets.  The absence of an extremely broad component in the high-$J$ CO and \water\ lines indicates fully dissociative shocks.  Tracing the ionized jet, the \feii\ emission from the ionized jet exhibits an extremely broad (FWHM $=150-180$ \kms) component centered at $\sim-$120 \kms\ and a narrower (FWHM $=40$ \kms) component at $-300$ to $-400$ \kms\ \citep{2009ApJ...694..654P}.  
While our observations only show the far-infrared \oi\ line up to $\sim-100$ \kms, the optical \oi\ line at 6364 \AA\ has a broad (FWHM $\sim$200 \kms) line at $\sim-100$ to $-$300 \kms\ similar to that characteristics of the \feii\ emission \citep{2004ApJ...616..998W}, suggesting that both \oi\ and \feii\ trace the jet.  The brightness of this extremely high-velocity component likely falls below the sensitivity of our observations.

\begin{figure}[htbp!]
  \centering
  \includegraphics[width=0.48\textwidth]{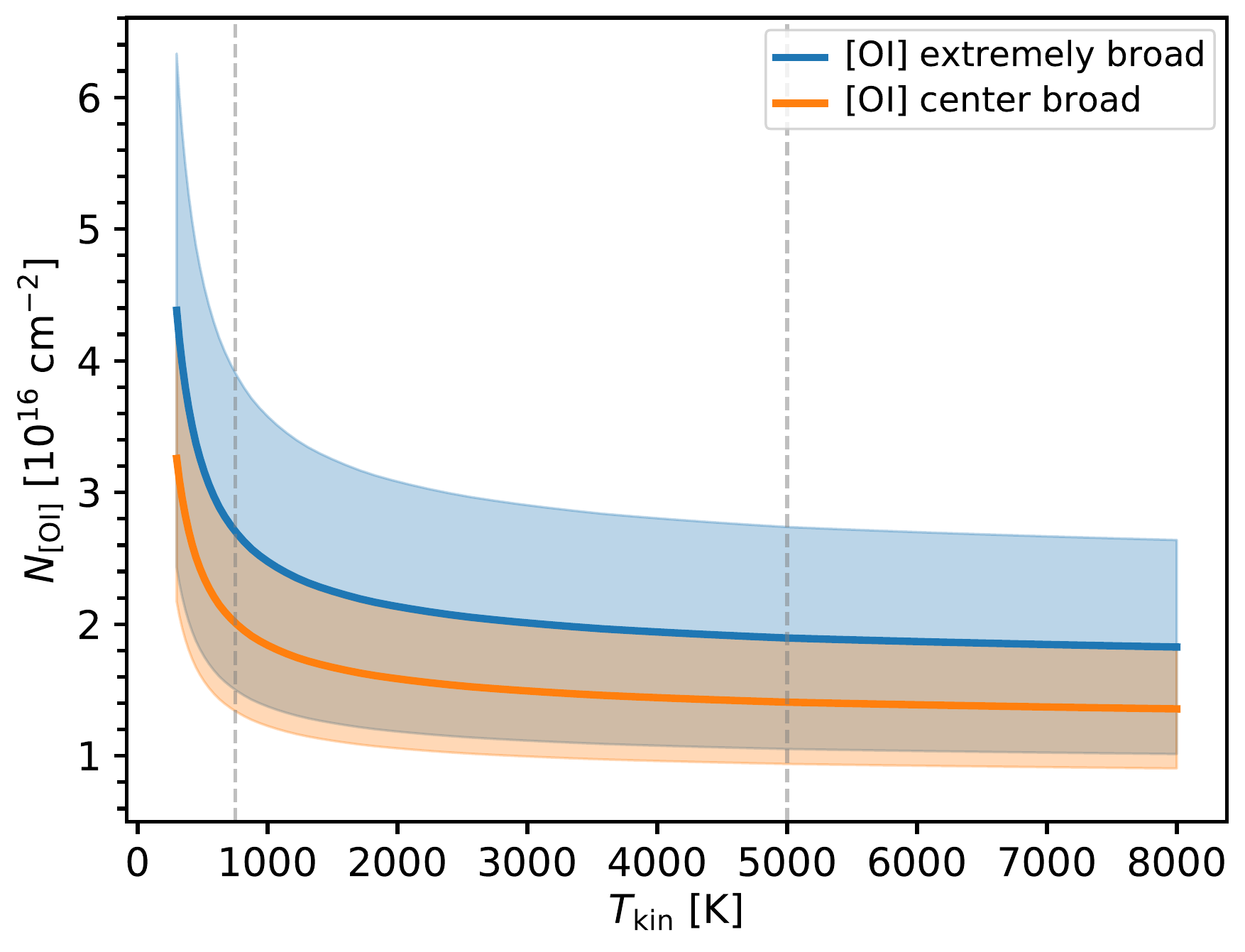}
  \caption{The best-fitting \oi\ column density from the \textsc{Radex} non-LTE modeling.  The orange and blue lines show the fitting results for the broad component at the source velocity and the extremely broad component, respectively.  The shaded region represents the uncertainty due to the noise in the spectra.  The vertical dashed lines highlight $T_\text{kin}=750$ K and 5000 K, relevant for cavity and J-shocks in the broad component at the source velocity and the extremely broad component, respectively.}
  \label{fig:oi_nlte}
\end{figure}

We can further estimate the oxygen budget in the shocks.  To estimate the \oi\ column density for the broad component centered at the source velocity (Figure\,\ref{fig:oi_nlte}, orange line), we used the non-LTE radiative transfer code, \textsc{Radex} \citep{2007AA...468..627V} and the collision rates from Leiden Atomic and Molecular Database \citep[LAMDA][]{2005AA...432..369S}.  While the ratio of \oi\ 63 \micron\ and 145 \micron\ line fluxes indicates a H$_2$ density of \eten{5.5} \cc\ \citep{2015ApJ...801..121N}, both the kinetic temperature ($T_\text{kin}$) and the column density of \oi\ remains unknown.  Using the line fitting results listed in Table\,\ref{tbl:fitting}, we modeled an \oi\ column density of (1--5)\ee{16} cm$^{-2}$ with a kinetic temperature ranging from 500 to 8000 K (Figure\,\ref{fig:oi_nlte}).  The modeling used the \oi\ atomic data from the Leiden Atomic and Molecular Database \citep{2005AA...432..369S} and assumed a uniform geometry for calculating the escape probability to reproduce the observed \oi\ line flux within 1\%.
The collision rates from LAMDA only cover the first three states of \oi.  To test the necessity of including higher states that produce the 6300 \AA\ and 6364 \AA\ lines, we further include the $^1$D$_2$ and $^1$S$_0$ states using Table F.3 and Equation 2.27 of \citet{draine2010physics}, which employs the data from \citet{1990AA...231..499P}.  Assuming a typical electron density of 2\ee{4} \cc\ in protostellar jets \citep{2019ApJ...876..147H}, the calculations with the updated collision rates show that these excited states have negligible impact on the atomic oxygen column density derived from the $^3$P$_1\rightarrow^3$P$_2$ transition.  The uncertainty of the \oi\ column density is $\sim40$\%\ due to the uncertainty in the measured flux. The modeled \oi\ column density only varies by a factor of $\sim$2 in $8000\,\text{K} >T_\text{kin} > 750$ K.  Because both atomic and molecular lines are present in the cavity shocks, the temperature has to be moderate to prevent dissociation.  Therefore, we assume that the cavity shocks in \source\ have the same temperature as the shocks in NGC 1333 IRAS 4A R1, 750 K \citep{2017AA...601L...4K}, which then gives a \oi\ column density of (2.0$\pm$0.7)\ee{16} cm$^{-2}$.
Assuming a temperature of 750 K and a H$_2$ density of \eten{5.5} \cc, we can also model the broad component of the CO and \water\ emission using \textsc{Radex}.  For the CO\,\jj{16}{15} line, the broad (FWHM $=14.9$ \kms) component has a peak intensity of 0.36 K (Figure\,\ref{fig:oi_fitting}), resulting in a column density of (2.7$\pm$0.2)\ee{15} cm$^{-2}$.  For \water, we modeled the red- and blue-shifted components separately.  The modeling suggests a column density of (5.1$\pm$2.6)\ee{12} cm$^{-2}$ for the red-shifted component in the $1_{11}\rightarrow0_{00}$ line with a FWHM of 26.1 \kms\ and a peak intensity of 0.02 K; the modeling meanwhile gives a column density of (5.0$\pm$2.3)\ee{12} cm$^{-2}$ for the blue-shifted component in the $2_{02}\rightarrow1_{11}$ line with a FWHM of 8.4 \kms\ and a peak intensity of 0.03 K.  

To derive the abundance of dominant oxygen carriers, we need to estimate the mass of each species in the same location.  Because of the different beam sizes for \oi, CO, and \water\ observations, we examine two scenarios to derive the range of possible abundances.  First, we assume that the shocks are spatially unresolved for all three lines.  In this case, we can derive the mass of each species by multiplying the column density with the respective beam size.  The SOFIA beam for \oi\ is 6\farcs{3} and the Herschel beams for CO\,\jj{16}{15} and \water\ lines are 11\farcs{5} and 22\arcsec, respectively.  The relative abundance of each species can be written as
\begin{equation}
  \frac{X({\rm Y})}{X(O_\text{total})} = \frac{N(Y)\Omega_\text{Y}}{N({\rm O})\Omega_\text{O} + N({\rm CO})\Omega_\text{CO} + N({\rm H_2O})\Omega_{\rm H_2O}
  },
  \label{eq:abundance}
\end{equation}
where $\Omega_{\rm Y}$ is the beam area for the observations of each species. 
In the second scenario, we assume the shocks are uniformly spread out over the beam.  A larger beam would probe more shocked gas.  Therefore, we derive the abundance of each species using the smallest beam size (i.e., 6\farcs{3} for \oi) instead of their respective beams.  In both cases, we assume that all three species trace the same shocks.  Furthermore, we only consider the gaseous oxygen carriers and the refractory oxygen remains locked on dust grains because of the SiO non-detection \citep{1999AA...343..585C}.  Any refractory oxygen released from dust grains will make the relative abundances in Table 3 upper limits.

For cavity shocks and/or disk winds, atomic oxygen is the primary carrier of oxygen, representing 69\%\ and 88\%\ of the gaseous oxygen abundance in the extended and unresolved scenarios, respectively, with CO as the secondary oxygen carrier (Table\,\ref{tbl:abundance}).

\begin{table*}[htbp!]
  \centering
  \caption{Estimated abundances of O, CO, and H$_2$O \label{tbl:abundance}}
  \begin{tabular}{cccc}
    \toprule
    Scenario   & $X({\rm O})/X(O_\text{total})$ & $X({\rm CO})/X(O_\text{total})$ & $X({\rm H_2O})/X(O_\text{total})$ \\
    \midrule
    Extended   & (6.9$\pm$2.4)\ee{-1} & (3.1$\pm$0.3)\ee{-1} & (2.1$\pm$1.0)\ee{-3} \\
    Unresolved & (8.8$\pm$3.1)\ee{-1} & (1.2$\pm$0.1)\ee{-1} & (2.0$\pm$1.0)\ee{-4} \\
    \midrule
               & $X({\rm O})^{a}$ & $X({\rm CO})^{a}$ & $X({\rm H_2O})^{a}$ \\
    \midrule
    Extended   & (1.4$\pm$0.5)\ee{-4} & (6.2$\pm$0.6)\ee{-5} & (4.2$\pm$2.0)\ee{-7} \\
    Unresolved & (1.7$\pm$0.6)\ee{-4} & (2.4$\pm$0.2)\ee{-5} & (4.0$\pm$2.0)\ee{-8} \\
    \bottomrule
    \multicolumn{4}{l}{$^{a}$The total volatile oxygen abundance to H ($X(O_\text{total})$) is assumed as 2\ee{-4}.}
  \end{tabular}
\end{table*}

To estimate the absolute abundances of these major oxygen carriers, we need to make assumptions about the total gaseous oxygen abundance.  There is a long-standing problem of missing oxygen, where the elemental oxygen abundance in the interstellar medium, 5.8\ee{-4} \citep{2008ApJ...688L.103P}, is consistently higher than all the observable oxygen carriers in star-forming regions \citep[e.g., ][]{2021AA...648A..24V}.  This so-called undetected depleted oxygen (UDO) abundance with respect to H is $\sim$1-2\ee{-4} in star-forming regions \citep{2021AA...648A..24V}.  If we assume a UDO abundance of 2\ee{-4}, the remaining volatile oxygen abundance is 3.8\ee{-4}, similar to the value of (3.4$\pm$0.5)\ee{-4} measured from using translucent sight lines toward stars \citep{2004ApJ...605..272S}.  Furthermore, refractory dust also carries oxygen.  Following the assumption made in \citet{2021AA...648A..24V}, we assume a refractory abundance of 1.4\ee{-4} derived from the Mg, Si, and Fe abundances in \citet{2008ApJ...688L.103P}.  Henceforth, we assume a total gaseous oxygen abundance of 2.0\ee{-4}.  From Equation\,\ref{eq:abundance}, we then derive an atomic oxygen abundance relative to H 
for the two scenarios described above (``extended'' and ``unresolved'' in   Table\,\ref{tbl:abundance}).  The dominant presence of \oi\ is expected in \source; \citet{2018ApJS..235...30K} found the \oi\ luminosity dominates the line luminosity of major coolants, such as \water, CO, and OH.  In \source, the \oi\ luminosity takes up 55\%\ of the total line luminosity, while the warm and hot CO luminosity takes up 38\%.  The $X({\rm O})$ is shockingly high compared to the oxygen abundance in other observations of protostellar shocks.  In the R1 shock knot of NGC 1333 IRAS 4A, for instance, \citet{2017AA...601L...4K} analyzed the velocity-resolved \oi\ emission and estimated an $X({\rm O})$ of 5--6\ee{-5}, taking up only $\sim$15\%\ of the total oxygen volatile abundance.

\subsection{Photodissociation of \water}

\begin{figure}[htbp!]
  \includegraphics[width=0.48\textwidth]{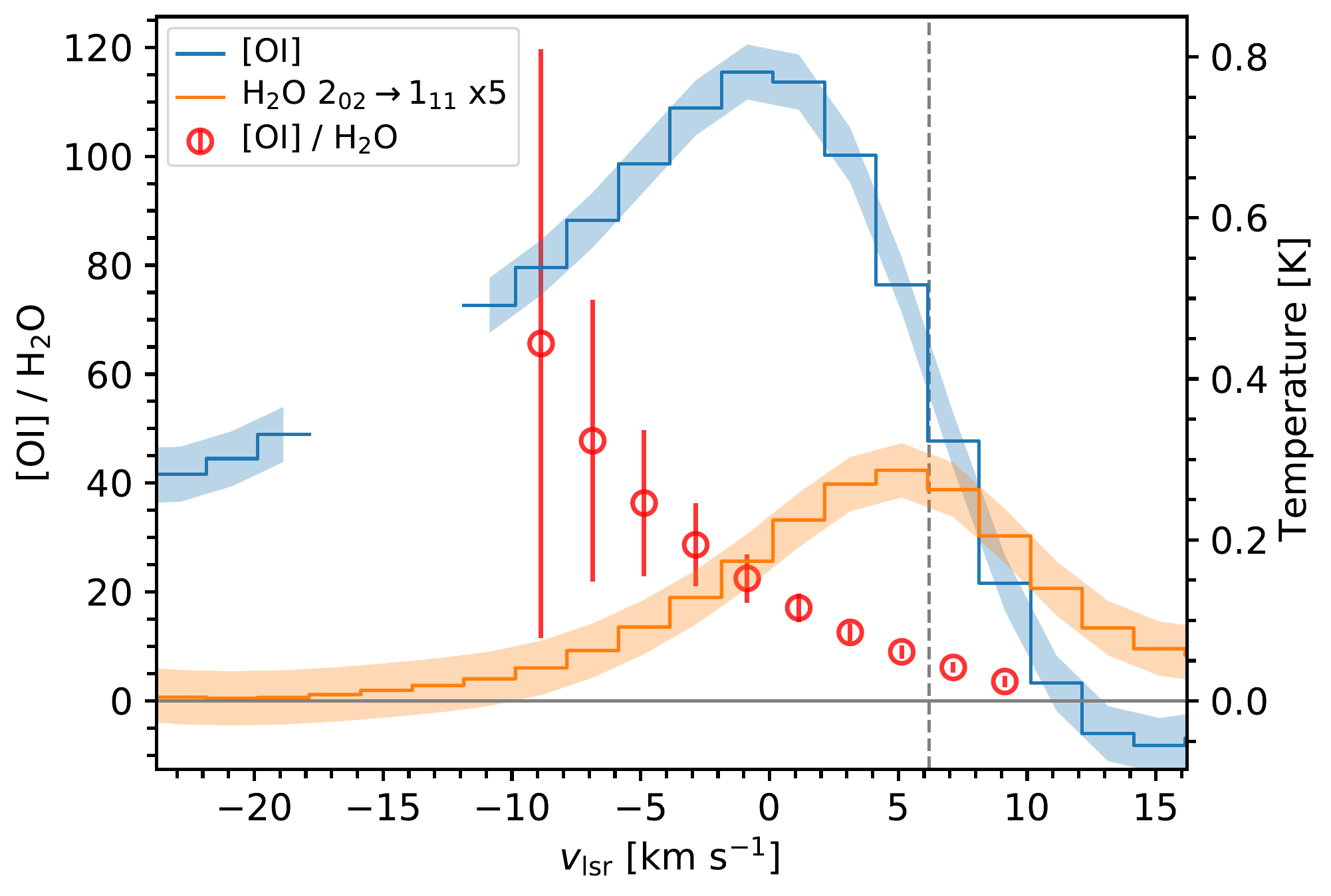}
  \caption{The ratio of \oi\ 63 \micron\ and the \water\ $2_{02}\rightarrow1_{11}$ line intensities as a function of velocity (red circles).  The spectra of \oi\ and \water\ are smoothed to 2 \kms\ before calculating the ratio.  Only data with a $>1\sigma$ signal is included when calculating the ratio.  The smoothed spectra of \oi\ and \water\ are shown (right axis) for comparison.  The shaded region shows the uncertainties of the \oi\ and \water\ smooth spectra in their respective colors.  The dashed vertical line indicates the source velocity, while the solid horizontal line highlights the baseline.}
  \label{fig:oi_h2o_ratio}
\end{figure}

The ratio of the \oi\ to \water\ emission can constrain the velocity where \water\ becomes dissociated.  For example, in irradiated C-shock models, the O abundance decreases with shock velocity, while the \water\ abundance increases, resulting in a decreasing \oi/\water\ ratio \citep{2015ApJ...806..227M,2019AA...622A.100G}, the opposite to this SOFIA observation shows.  A high UV radiation would promotes photodissociation, efficiently destroying \water; however, \citet{2020AA...643A.101L} find only J-shocks can produce sufficient UV radiation to generate significant photodissociation.  Smooth, steady-state disk winds can reproduce the Herschel \water\ observations that trace outflows and contain little dust to block the UV radiation due from the accreting protostar \citep{2016AA...585A..74Y}.  Thus, the cavity shock component can be a C-type shock irradiated by an exceptional amount of UV radiation from sources other than the shock itself, a J-type shock that shows increasing O abundance with velocities \citep{2020AA...643A.101L}, or disk winds. In Figure\,\ref{fig:oi_h2o_ratio}, we show the \oi/\water\ intensity ratio as a function of velocity.  The ratio has a minimum at $\sim$9 \kms, a value of $\sim$5 at the source velocity, 6.2 \kms, and then increases to $\sim$60 at $-10$\kms.

Only the \oi\ emission shows an extremely broad component, suggesting that the shocks are fully dissociative.  The modeled column density of this component is (2.7$\pm$1.6)\ee{16} cm$^{-2}$, assuming a kinetic temperature of 750 K.  The models of FUV irradiated C-shocks break down at velocity $\sim$20 km s$^{-1}$, assuming a pre-shock density of $10^{5}$ cm$^{-3}$, becoming dissociative at larger velocities \citep{2015ApJ...806..227M}. The lack of \water\ emission beyond $v\lesssim-15$ \kms\ is consistent with its photodissociation in a J-type shock.  \citet{1985Icar...61...36H} presented a simple J-shock model, suggesting that \oi\ emission dominates the cooling at $T \lesssim 5000$ K.  If we take $T = 5000$ K as an upper limit of the \oi\ temperature for the extremely broad component, we can estimate a lower limit of \oi\ column density of 2\ee{16} cm$^{-2}$ from the non-LTE radiative transfer modeling (Figure\,\ref{fig:oi_nlte}). 

\subsection{Mass Loss Rates}

A longstanding issue with molecular outflows is whether they can be driven by stellar winds or jets. The basic idea is that the wind emerges from the star and/or disk at high speeds and sweeps out a cavity in the infalling core. The swept-up gas is observed in low rotational transitions of CO. The highly radiative nature of the shocks in the molecular gas imply that only the momentum of the wind is available to sweep up material (so-called ``momentum conserving'' situations). In this picture, the momentum in the CO flow must be  matched by the momentum of the wind,
\begin{equation}
P_\text{w} = P_{\rm CO} = M_\text{w} v_\text{w} = M_{\rm CO} v_{\rm CO}
\label{eq:momentum}
\end{equation}
If this is correct, an intrinsic mass loss rate in the wind can be derived from
\begin{equation}
\mdotstar = P_{\rm CO}/(t_{\rm CO} v_\text{w}) = F_{\rm CO}/v_\text{w},
\label{eq:mdot}
\end{equation}
where $F_{\rm CO}$ is the average force that must have been applied to the swept-up gas, $v_{\rm CO}$ is the velocity of the swept-up gas, and $v_\text{w}$ is the velocity of the stellar wind.  The intrinsic mass loss rate is averaged over the age of the flow, $t_{\rm CO}$. This basic idea was developed by \citet{1988ApJS...67..283L} and applied to CO maps of young stellar objects, including \source.  Maps of CO are readily obtained, while observations of the wind itself have proved challenging. Since the latter inevitably average over much shorter times, comparison of the two provides a potential means to measure changes, whether secular or episodic, in the intrinsic mass loss rate (\mdotstar), which in turn, is likely to be closely related to the mass accretion rate onto the star, a quantity of considerable interest.  The binary protostars of \source\ drive two jets, as can be seen in \feii\ \citep{1998ApJ...499L..75F,2009ApJ...694..654P}, but remain unresolved with \oi\ and other tracers.  Thus, we are measuring the total intrinsic mass loss rate instead of rates for individual jets.

There have been many maps of different transitions of CO toward \source, with a wide range of inferred momenta. For the most stringent test of the constancy of the mass loss rate, we prefer the largest, most complete maps of the lower $J$ transitions of CO, which turn out to be the oldest.  The outflow was the first bipolar flow discovered \citep{1980ApJ...239L..17S}, but
better maps were made in the \jj{1}{0} transition by \citet{1985ApJ...295..490S} and the \jj{2}{1} transition by \citet{1988ApJS...67..283L} both covering about 30\arcmin.

Determination of the force from the observations can take various paths, with issues of how to handle the fact that the outflow is inclined to the plane of sky ($\theta_\text{incl}$). Some definitions of the inclination angle use the angle between the outflow axis and the plane of the sky, while others use the line of sight to the observer as the reference. We adopt the former.  The \source\ outflow is mostly in the plane of the sky, and we adopt $\theta_\text{incl}=30\arcdeg$, based on a determination of the disk angle (90-$\theta_\text{incl}$) of $60\arcdeg\pm5$\arcdeg\ \citep{2014ApJ...796...70C}.  

Following \citet{2014ApJ...783...29D}, we make no correction to the observed $P_{\rm CO}$ (see also \citealt{2007AA...471..873D}) but correct $t_{\rm CO}$ because both length and velocity are affected.  For our convention on inclination angle,
\begin{equation}
t_{\rm CO} = t_{\rm obs} \times {\rm tan} (\theta_\text{incl}).
\label{eq:age_correction}
\end{equation}
This approach is similar to Method M7 of \citet{2013AA...556A..76V}.  Therefore, forces determined from observations, with no corrections, should be multiplied by $1/{\rm tan}(\theta_\text{incl}) = 1.73$. We added forces for the blue and red lobes. The result for the \jj{1}{0} map of \citet{1985ApJ...295..490S}, after correction of age by inclination angle, is $F_{\rm CO} = 1.2\ee{-4}$ \msunyr\kms\ averaged over an age of 4.6\ee{4} yr. The result from the \jj{2}{1} map from \citet{1988ApJS...67..283L}, is $F_{\rm CO} = 2.1\ee{-4}$ \msunyr\kms\ averaged over an age of 2.9\ee{4} yr corrected by inclination angle.

The wind velocity has been estimated from spectra of \feii\ $\lambda = 1.644$ \micron\ emission.  \citet{2005ApJ...618..817P} find velocities as high as 300 \kms, with a blue-shifted components at $100$ \kms\ to $279\pm 10$ \kms.  \citet{1992ApJ...397..214G} also observed velocities up to 150 \kms\ in HI.  Correcting for inclination by $1/{\rm sin} (\theta_\text{incl}) = 2$ implies $v_\text{w} = 200$ to $560$ \kms\ for ionized gas and $v_\text{w} = 260$ \kms\ for the neutral gas.  Such high velocities strongly indicate an origin for the wind near the star, rather than a disk wind.  The SOFIA \oi\ observations only detect \oi\ up to $\sim$75 \kms\ in blue-shift, which may be due to insufficient sensitivity.  We used $v_\text{w}=300$ \kms\ as a compromise.  The values for \mdotstar\ are then 4.0\ee{-7} \msunyr\ for \jj{1}{0} and 7.0\ee{-7} \msunyr\ for \jj{2}{1}.

\citet{2015AA...576A.109Y} also mapped the outflow of \source\ using the CO\,\jj{3}{2} line to derive the force and dynamical age of the blue- and red-shifted lobes; however, the map size is only $\sim5\arcmin\times5\arcmin$.  We use their Equation 4 to retrieve the momentum without any correction, 0.11 and 0.40 \msun\ \kms, for the blue- and red-shifted outflow, respectively.  They also measured a dynamical age of 8.3\ee{3} and 6.7\ee{3} yr uncorrected for inclination in the blue- and red-shifted outflow, respectively.  Using Equation\,\ref{eq:mdot} to correct the averaged age for inclination angle, and $v_\text{w}=300$ \kms, we derive an intrinsic mass loss rate of 4.0\ee{-7} \msunyr, consistent with the estimates from low-$J$ CO emission.

In the momentum-conserving scenario, atomic winds are thought to drive the molecular outflows because the ionized wind/jet has too little momentum \citep{1987ApJ...319..275M}.  \citet{1992ApJ...397..214G} observed the high velocity wings of HI line profiles using the Arecibo Observatory.  Assuming optically thin emission and mass conservation of HI, they modeled the line profiles with a decelerating wind model to get $\mdotstar= 2.4\ee{-6}$ \msunyr.  However, their model assumes that all HI gas is due to the wind, while photodissociation in the envelope can produce HI.  Moreover, the HI spectra suffer from intervening structured clouds at the blue-shifted velocity.  Thus, our SOFIA \oi\ observations would better measure \mdotstar\ in atomic winds and so, in the following discussion, we focus on the extremely broad component of \oi.

\oi\ is a good tracer of atomic outflows in the model proposed by \citet{1985Icar...61...36H}.  The model has a fast ($v_{\rm w}$) wind from the star-disk system driving a shock into the infalling molecular gas, pushing it away from the forming star at speeds of a few to tens of \kms\ ($v_{\rm m}$).  The slowing of the wind in the molecular shock then drives a reverse shock into the wind at $v_{\rm wsh} = v_{\rm w} - v_{\rm m} \sim v_{\rm w}$ \kms.  In this model, the molecular shock is formed in the ambient material, producing the emission from CO, H$_2$, H$_2$O, and OH.  The rapid deceleration of the wind as it encounters the molecular material produces the wind shocks, resulting in a J-type shocked region, with $T_{\rm K} = 10^3-10^5$ K.  Such a high temperature would dissociate most molecules, making the fine-structure \oi\ 63 \micron\ line the dominant coolant in the shocked wind.  The shocks are strongly radiative and only the momentum of the wind is assumed to be available to accelerate the molecular outflow.  Thus, in this model, \oi\ 63 \micron\ emission can conveniently be related to the intrinsic mass loss rate by \citet{1985Icar...61...36H}
\begin{equation}
  \frac{L_{{\rm [OI]}\,63\,\mu m}}{L_{\odot}} = \frac{0.1 \mdotstar}{10^{-5} M_{\odot}\,\text{yr}^{-1}}.
  \label{eq:mdot_Loi}
\end{equation}
Using the flux of the extremely broad \oi\ component that traces high-velocity gas (Table\,\ref{tbl:fitting}), we can derive $\mdotstar=3.4\pm1.8\ee{-7}$ \msunyr\ using the observed \oi\ luminosity, 3.4$\pm$1.8\ee{-3} \lsun, assuming that the missing red-shifted \oi\ emission has the same luminosity as the observed blue-shifted part.  If we adopt the total \oi\ 63 \micron\ luminosity from Herschel observations, 4.8$\pm$0.1\ee{-3} \lsun\ \citep{2016AJ....151...75G}, \mdotstar\ becomes 4.8$\pm$0.1\ee{-7} \msunyr.  \citet{2021AA...650A.173S} also estimated \mdotstar\ using the \oi\ emission observed with SOFIA/FIFI-LS.  They employed two methods, one assuming collisionally excited \oi\ emission from purely atomic oxygen gas \citep{2020AA...642A.216S} and the other one using the relation in \citet{1985Icar...61...36H}, deriving \mdotstar\ of 5.8--11.8\ee{-7} and 4.9--5.4\ee{-7} \msunyr, respectively.  

Equation\,\ref{eq:mdot_Loi} provides a zeroth-order estimate of the intrinsic mass loss rate and there are important caveats.  This method assumes the entire wind shock radiates before any gas gets shock-heated or entrained, which only occurs when measuring the jet close to its source \citep{2019ApJ...876..147H}.  Secondly, if the observations probe several unresolved shocks instead of a single shock, Equation \ref{eq:mdot_Loi} may overestimate the intrinsic mass loss rate \citep{2015ApJ...801..121N}.  Finally, if parts of the wind propagate without intervention, this method would underestimate the intrinsic mass loss rate \citep{1988ApJ...329..863C}.
With the velocity-resolved SOFIA \oi\ observations, we can directly estimate the mass outflow rate traced by \oi\ to partially circumvent these caveats, especially the first one.  If the extremely broad component of \oi\ traces the atomic wind that carries most momentum in the wind, we can simply estimate \mdotstar\ by taking the ratio of the mass traced by \oi\ and the dynamical time.  With a column density of 2.7$\pm$1.6\ee{16} cm$^{-2}$, we can estimate the mass outflow rate using
\begin{align}
    \mdotstar & = \frac{\mu m_\text{H} N_\text{[OI]} \Omega_\text{b}/X_\text{O}}{t_\text{dyn}}; \nonumber \\
  t_\text{dyn} & = \frac{R_\text{beam}}{v_\text{[OI]}}\,\text{tan}(\theta_\text{incl}),
  \label{eq:mdot_oi}
\end{align}
where $\mu$ is the mean molecular weight of 2.8 \citep{2008AA...487..993K}, $\Omega_\text{b}$ (1.13$\times$6\farcs{3}$^2$) is the area of the Gaussian beam with FWHM=6\farcs{3}, $X_\text{O}$ is the atomic oxygen abundance, $t_\text{dyn}$ is the inclination-corrected dynamical time of the \oi\ gas, $R_\text{beam}$ is the radius of the beam, and $v_\text{[OI]}$ is the velocity of \oi\ gas.  We estimate the dynamical age of the wind as the time required for the gas to cross the beam radius with correction of inclination angle.  The $t_\text{dyn}$ is 35 yr.  The atomic oxygen abundance is a key parameter to relate the \oi\ mass flux to the intrinsic mass outflow rate.  Because neither the \water\ nor high-$J$ CO emission shows an extremely broad component, this component is likely to be fully dissociated.  Thus, if we assume a total oxygen volatile abundance ($X_\text{O}$), 3.4\ee{-4}, as discussed in Section\,\ref{sec:abundance}, and an equal rate in each outflow lobe, the intrinsic mass outflow rate becomes 1.3$\pm$0.8\ee{-6} \msunyr.  Considering the uncertainty, the intrinsic mass outflow rate from \oi\ agrees with not only the estimate from HI but also our estimate from the CO outflows after inclination correction, unambiguously confirming the momentum-driven outflows.  In this scenario, both atomic and ionized wind belongs to the mass ejected near from the protostar but the atomic wind carries most of the momentum, driving the molecular outflows.  Given its spectral profile in the optical and near-infrared, some studies consider \source\ as a FU Orionis-like object \citep[e.g.,][]{1985ApJ...297L..41M,2010AJ....140.1214C}, which implies it should have a strongly varying time-dependent mass loss rate.  But the similar \mdotstar\ found over a 3--5\ee{4} yr (low-$J$ CO) and 35 yr (\oi) timescales suggests no evidence of varying \mdotstar\ within a factor of 3.

\section{Conclusions}
\label{sec:conclusions}
This work presents SOFIA/upGREAT observations of \oi\ and \cii\ emission toward a Class I protostar, \source.  The main conclusions are listed below.

\begin{itemize}
  \item The SOFIA/upGREAT observations show an asymmetric profile for the \oi\ 63 \micron\ line toward the center of \source.  Emission of \oi\ only appears in the blue-shifted velocity with a FWHM of $\sim$100 \kms.  A two-component model can reproduce this \oi\ spectrum, indicating a $\sim$20 \kms\ width component at the source velocity and another extremely wide component, 87.5 \kms, at $-$30 \kms.
  \item The \oi\ line profile is consistent with the scenario of cavity shocks and spot shocks due to the outflow-envelope interaction.  The \oi\ component centered at the source velocity appears similar to the line profiles of high-$J$ CO and \water\ lines, which have a $\sim$20 \kms\ component around the source velocity tracing the cavity shocks.  Spot shocks would produce a line profile similar to the velocity-offset extremely broad \oi\ component.  This extremely broad component is only detected in \oi\ emission.  The non-detection of a narrow ($<5$ \kms) \oi\ component suggests that shocked gas dominates the \oi\ emission.
  \item Narrow \cii\ emission is detected at the protostar and in the red-shifted outflow with a S/N of 4.1 and 3.4, and line widths of 2.3 and 2.8 \kms, respectively.  The narrow line width suggests a PDR origin for the \cii\ emission.  Comparing to the Herschel observations, we estimate the PDR contribution to \oi\ and \cii\ line flux is $\sim$3\%\ and $\sim$30\%, respectively.
  \item We use non-LTE radiative transfer calculations to derive the column densities of \oi, CO, and \water\ in the component tracing the cavity shocks.  We consider the scenarios that the emission is uniformly extended and unresolved.  The atomic oxygen abundance is 69$\pm$24\%\ and 88$\pm$31\%\ of the total oxygen volatile abundance in \source\ for the extended and unresolved scenarios, respectively.  Assuming an observable oxygen abundance of 3.4\ee{-4} and a refractory dust abundance of 1.4\ee{-4}, we estimate an atomic oxygen abundance of 1.4$\pm$0.5\ee{-4} and 1.7$\pm$0.6\ee{-4} with respect to atomic H for the extended and unresolved cases, respectively, indicating that the emission originates in atomic cavity shocks.
  \item The intrinsic mass loss rate (\mdotstar) derived from \oi\ component is 1.3$\pm$0.8\ee{-6} \msunyr.  Observations of CO outflows, HI wind, and \oi\ wind produce consistent values of \mdotstar, unambiguously confirming the momentum-conserving scenario of outflows.  There is no evidence of varying \mdotstar\ over 3--5\ee{4} yr to a factor of 3.
\end{itemize}

These SOFIA/upGREAT observations suggest a highly atomic outflow in \source, opposite to the conclusions of previous \oi\ observations in the outflow of low-mass protostars \citep{2017AA...601L...4K}.  With velocity-resolved spectra, we can directly identify components of shocked gas and investigate their physical origin.  Future observations with high sensitivity can reduce the uncertainties presented in this study.  Moreover, velocity-resolved \oi\ emission has only previously been observed toward two protostellar outflows.  Thus, a full picture of the outflow-envelope interaction requires future studies on the shocked gas in outflows as well as more \oi\ observations in outflows.

\acknowledgements
The authors thank Edward Chambers and Simon Coud\'{e} for the observing support.  Y.-L.\ Yang acknowledges the support from the Virginia Initiative of Cosmic Origins (VICO) Postdoctoral Fellowship.  Y.-L.\ Yang also appreciates discussions with Fernando Cruz-Saenz de Miera, \'{A}gnes K\'{o}sp\'{a}l, Michihiro Takami, and Ilse Cleeves.  This study is based on observations made with the NASA/DLR Stratospheric Observatory for Infrared Astronomy (SOFIA). SOFIA is jointly operated by the Universities Space Research Association, Inc.\ (USRA), under NASA contract NNA17BF53C, and the Deutsches SOFIA Institut (DSI) under DLR contract 50 OK 0901 to the University of Stuttgart. Financial support for this work was provided by NASA through award \#NAS2-97001, SOF 04-0146 and 06-0104, issued by USRA.  A.K.\ acknowledges support from the First TEAM grant of the Foundation for Polish Science No.\ POIR.04.04.00-00-5D21/18-00. This article has been supported by the Polish National Agency for Academic Exchange under Grant No.\ PPI/APM/2018/1/00036/U/001. J.P.R.\ acknowledges support from the Virginia Initiative on Cosmic Origins (VICO), the National Science Foundation (NSF) under grant nos. AST-1910106 and AST-1910675, and NASA via the Astrophysics Theory Program under grant no. 80NSSC20K0533. Herschel is an ESA space observatory with science instruments provided by European-led Principal Investigator consortia and with important participation from NASA.

\facilities{\textit{SOFIA}}

\software{Class v. sep20a \citep{2013ascl.soft05010G}, Astropy v. 4.2 \citep{2013AA...558A..33A,2018AJ....156..123A}, Matplotlib v. 3.3.4 \citep{Hunter:2007}, JupyterLab v. 3.0.16 \citep{Kluyver2016jupyter}}


\end{document}